\documentclass[fleqn,usenatbib]{mnras}

\usepackage{newtxtext,newtxmath}

\usepackage[T1]{fontenc}
\usepackage{ae,aecompl}

\usepackage{booktabs}
\usepackage{multirow}
\usepackage{pdflscape}
\usepackage{float}
\usepackage[labelfont=bf]{caption}
\usepackage{comment}
\usepackage{graphicx}	
\usepackage{amsmath}	
\usepackage{amssymb}	

\newcommand{\comm}[1]{}

\title[K2 TFAW survey. I]{TFAW survey. I. Wavelet-based denoising of K2 light curves. Discovery and validation of two new Earth-sized planets in K2 campaign 1}

\author[D. del Ser et al.]{
D. del Ser,$^{1,2,3}$\thanks{E-mail: dser@fqa.ub.edu (DdS)}
and O. Fors$^{1,2,3}$
\\
$^{1}$Dept. de F{\'i}sica Qu{\`a}ntica i Astrof{\'i}sica, Institut de Ci{\`e}ncies del Cosmos (ICCUB), Universitat de Barcelona, IEEC-UB, Mart\'{\i} i \\
Franqu{\`e}s 1, E08028 Barcelona, Spain\\
$^{2}$Observatori Fabra, Reial Acad{\`e}mia de Ci{\`e}ncies i Arts de Barcelona, Rambla dels Estudis, 115, E-08002 Barcelona, Spain\\
$^{3}$University of North Carolina at Chapel Hill, Department of Physics and Astronomy, Chapel Hill, NC 27599-3255, USA\\
}

\date{Accepted 2020 August 14. Received 2020 August 11; in original form 2020 April 3}

\pubyear{2020}

\begin{document}
\label{firstpage}
\pagerange{\pageref{firstpage}--\pageref{lastpage}}
\maketitle

\begin{abstract}
 The wavelet-based detrending and denoising method \texttt{TFAW} is applied for the first time to \texttt{EVEREST 2.0}-corrected light curves to further improve the photometric precision of almost all K2 observing campaigns (C1-C8, C12-C18). The performance of both methods is evaluated in terms of 6 hr combined differential photometric precision (CDPP), simulated transit detection efficiency, and planet characterization in different SNR regimes. On average, \texttt{TFAW} median 6hr CDPP is $\sim$30$\%$ better than the one achieved by \texttt{EVEREST 2.0} for all observing campaigns. Using the \texttt{transit least-squares} (\texttt{TLS}) algorithm, we show that the transit detection efficiency for simulated Earth-Sun-like systems is $\sim$8.5$\times$ higher for \texttt{TFAW}-corrected light curves than for \texttt{EVEREST 2.0} ones. Using the light curves of two confirmed exoplanets, K2-44 b (high-SNR) and K2-298 b (low-SNR), we show that \texttt{TFAW} yields better MCMC posterior distributions, transit parameters compatible with the cataloged ones but with smaller uncertainties and narrows the credibility intervals. We use the combination of \texttt{TFAW}'s improved photometric precision and \texttt{TLS} enhancement of the signal detection efficiency for weak signals to search for new transit candidates in K2 observing campaign 1. We report the discovery of two new K2-C1 Earth-sized planets statistically validated, using the \texttt{vespa} software: EPIC 201170410.02, with a radius of 1.047$^{+0.276}_{-0.257}R_{\oplus}$ planet orbiting an M-type star, and EPIC 201757695.02, with a radius of 0.908$^{+0.059}_{-0.064}R_{\oplus}$ planet orbiting a K-type star. EPIC 201757695.02 is the 9-th smallest planet ever discovered in K2-C1, and the 39-th smallest in all K2 campaigns.
\end{abstract}




\section{Introduction}
\label{sec:intro}
Developed in the months following the failure of the second of the four reaction wheels of \emph{Kepler}, the K2 mission \citep{Howell2014}, represented a new concept for \emph{Kepler}'s operations given the spacecraft's ability to maintain pointing in all three axes with only two reaction wheels. This operation mode, that started in October 2013 and became fully operational in May 2014, provided enough fuel to began a series of 19 sequential campaigns observing a set of independent target fields distributed around the ecliptic plane. During this \emph{Kepler} extended mission, among other community-proposed targets, late-type dwarf stars were favored as targets due to the highest chance of detecting small planets lying in the habitable zone of their host stars. However, the failure of the reactions wheels degraded the photometric precision obtained for K2. Several decorrelation techniques were developed to improve the noise properties of the K2 light curves: pixel level decorrelation (PLD)~\citep{Deming2015}, \texttt{K2P} \citep{Lund2015}, \texttt{K2SFF} \citep{Vanderburg2014}, and \texttt{EVEREST 2.0}\footnote{\url{https://github.com/rodluger/everest}}~\citep{Luger2018}. The latter, based on a PLD combined with a Gaussian Process (GP) optimization led to best photometric precision to date achieved with K2 light curves.

In many instances the systematic variations in a given light curve are also shared by other stars and same data set. A common approach to remove those systematics is to identify the objects in the field that suffer from the same kind of variations as the target (e.g. correlated noise), and then apply a filter based on the light curves of those template stars. The Trend Filtering Algorithm (hereafter \texttt{TFA})~\citep{Kovacs2005} is often applied to remove systematic variations in ground-based time-domain surveys, in particular the ones searching exoplanetary transits and variable stars. 

Optimizing the photometric precision achieved by an astronomical survey is a key factor to increase the probability of detecting periodic signals in the data. Here we apply \texttt{TFAW}~\citep{delSer2018}, the wavelet-based modification of \texttt{TFA}, to further improve the photometric precision achieved by \texttt{EVEREST 2.0} data. \texttt{TFAW} uses the wavelet transform's signal decoupling and denoising potential to estimate the noise contribution of the light curve and the shape of the underlying signal and, returns a denoised signal without modifying any of its intrinsic properties. We combine this improved photometric precision achieved by \texttt{TFAW} with the optimized detection of small planets obtained by the \texttt{transit least-squares} (hereafter \texttt{TLS}) algorithm \citep{Hippke2019} to search for periodic signals, in particular, planetary transit ones, within the \texttt{TFAW}-corrected light curves.

In Section \ref{sec:tfaw} we briefly describe the \texttt{TFAW} algorithm. In Section \ref{sec:performance} we compare the \texttt{TFAW} performance, in terms of combined differential photometric precision, with respect to the one obtained with \texttt{TFA} and \texttt{EVEREST 2.0} K2 data. We also evaluate the transit detection efficiency of Earth-Sun-like injected systems, both in \texttt{EVEREST 2.0} and \texttt{TFAW} detrended light curves. In addition, the biases and uncertainties of the fitted transit parameters values for two known K2 planetary systems (K2-44 b and K2-298 b), are also compared for both \texttt{TFAW} and \texttt{EVEREST 2.0} detrended light curves. In Section \ref{sec:k2c1}, we present the transit search and vetting criteria employed with our \texttt{TFAW} light curves from K2 observing campaign 1, compare our transit search results with the ones from other groups and, finally, we present two new statistically validated Earth-sized candidate planets found using \texttt{TFAW}-corrected light curves and \texttt{TLS}.

\section{\texttt{TFAW} detrending algorithm}
\label{sec:tfaw}

\texttt{TFAW} \citep{delSer2018} combines the detrending and systematics removal capabilities of \texttt{TFA} with the signal decoupling and denoising potential of the wavelet transform. The noise contribution and the underlying signal shape are iteratively estimated from the target light curve using the Stationary Wavelet Transform (hereafter SWT). This allows \texttt{TFAW} to denoise and reconstruct the signal without modifying any of its astrophysical properties, i.e., without introducing artificial distortions in the signal's shape or ripples around discontinuities.

The \texttt{TFAW} detrending algorithm can be summarized as follows (see \citet{delSer2018} for further details): 1) an initial filter is computed, as with the original \texttt{TFA}, by means of a template of reference stars, to remove trends and systematics from the target light curve, 2) the signal shape of the detrended light curve is inferred by means of the SWT decomposition levels and its corresponding power spectrum. The outliers from the light curve are removed and a first estimation of the high frequency noise contribution is removed using the lowest SWT decomposition level (i.e. the one with highest frequency resolution), 3) frequency analysis step: a search for significant periodicities is run over this outlier-free and denoised signal, 4) if a significant periodicity is found, the shape of the trend- and noise-free phase folded signal is estimated again using the SWT, and 5) signal reconstruction step: the trend-free light curve is iteratively denoised and reconstructed during \texttt{TFAW} signal reconstruction process. 

The original \texttt{TFAW} implementation used the \texttt{BLS} algorithm \citep{Kovacs2002} to search for significant transit-like periodicities within the target light curves. We have extended the algorithm capabilities to detected transits of smaller planets by using the \texttt{TLS} algorithm that takes the stellar limb darkening and planetary ingress and egress into account. We consider a periodicity to be significant if it corresponds to the highest peak in the \texttt{TLS} power spectrum and its Signal Detection Efficiency (SDE$_{\rm TLS}$) value is above 9.0. Any light curve that matches this criteria during \texttt{TFAW}'s signal detection step will undergo the iterative signal reconstruction and denoising.

\section{\texttt{TFAW} vs. \texttt{EVEREST 2.0} performance}
\label{sec:performance}

In \citet{delSer2018} we show that \texttt{TFAW} improves the detection rate, denoising and characterization of different astrophysical periodic signals compared to \texttt{TFA}.
In this Section we want to assess \texttt{TFAW}'s performance when applied to non-median filtered (to avoid removing any stellar variability of interest) \texttt{EVEREST 2.0} light curves from the K2 mission. We do so by measuring the 6 hr combined differential photometric precision (CDPP) \citep{Christiansen2012,Luger2018} for \texttt{TFA}, \texttt{TFAW} and \texttt{EVEREST 2.0}, the transit detection efficiency obtained with \texttt{TFAW} and \texttt{EVEREST 2.0}, and comparing the biases and uncertainties of the Markov Chain Monte Carlo (MCMC) fitted transit parameters values for two known K2 planetary systems for \texttt{TFAW} denoised and reconstructed light curves and the \texttt{EVEREST 2.0} originals.

\subsection{Data selection}

To run \texttt{TFAW}, we download all the \texttt{EVEREST 2.0} light curves from the K2 mission monitoring campaigns C1 to C8, and C12 to C18 available at the MAST archive\footnote{\url{https://archive.stsci.edu/hlsps/everest/v2/bundles/}} earlier than 4 Jan 2019. We do not consider light curves from campaigns C9 (used to study gravitational microlensing events), and from C10 to C11 (they are both split in separate sub-campaigns due to a pointing and initial roll-angle error, respectively). For this work, we focus only on the long cadence (LC) light curves as the number of available template stars per CCD module is greater than for the short cadence (SC) data. Table~\ref{tab:cdpp_lcvs} lists the number of \texttt{EVEREST 2.0} light curves for each K2 campaign used in this work.

Given that the SWT needs an even number of data points to work, for campaigns C1-C8 we use 3072 epochs and 2432 for campaigns C12-C18. This way we ensure a good number SWT decomposition levels (10 and 7, respectively) to determine the signal and noise contributions of the light curves while keeping enough epochs to run the periodic signal search. As neither \texttt{TFA} or \texttt{TFAW} are designed to deal with Pixel Level Decorrelation (PLD) \citep{Deming2015}, we use the PLD, single co-trending basis vector (CBV), corrected fluxes provided by the \texttt{EVEREST 2.0} pipeline. For all light curves only epochs with the \texttt{QUALITY}=0 flag are considered (as described in \citet{Luger2018}) and extreme outliers are removed prior to the analysis. 

\begin{table}
\centering
\caption{Number of \texttt{EVEREST 2.0} long cadence light curves used by \texttt{TFA} and \texttt{TFAW} for different K2 campaigns. C9 is not considered. C10 and C11 are not used because both are split in separate sub-campaigns.} 
\label{tab:cdpp_lcvs} 
\begin{tabular}{cc}
\hline
K2 campaign & \texttt{EVEREST 2.0} light curves\\
\hline
C1  & 18703\\
C2  & 13394\\
C3  & 14151\\
C4  & 15539\\
C5  & 23074\\
C6  & 27435\\
C7  & 13483\\
C8  & 21387\\
C12 & 27524\\
C13 & 21407\\
C14 & 19230\\
C15 & 22814\\
C16 & 23506\\
C17 & 30931\\
C18 & 19053\\
\hline
\end{tabular}
\end{table}

\subsection{Combined differential photometric precision (CDPP)}
\label{subsect:cdpp_allcvs}

As a figure of merit to compare {\tt EVEREST 2.0}, \texttt{TFA} and \texttt{TFAW} photometric performance, we adopt the 6 hr CDPP. In practice we use the same CDPP metric as \texttt{EVEREST 2.0}, the one computed by smoothing the light curve with a 2-day, quadratic Savitsky-Golay filter, clipping outliers at 5$\sigma$, computing the median standard deviation in 13-cadence segments, and normalizing by $\sqrt{13}$. Though it might not be appropriate for very large amplitude or very short period variabilities, we choose the 6 hr timescale on purpose as it is roughly the transit duration of an Earth-Sun analogue \citep{Christiansen2012}.

\begin{figure*}
\centering
\includegraphics[height=18cm,keepaspectratio]{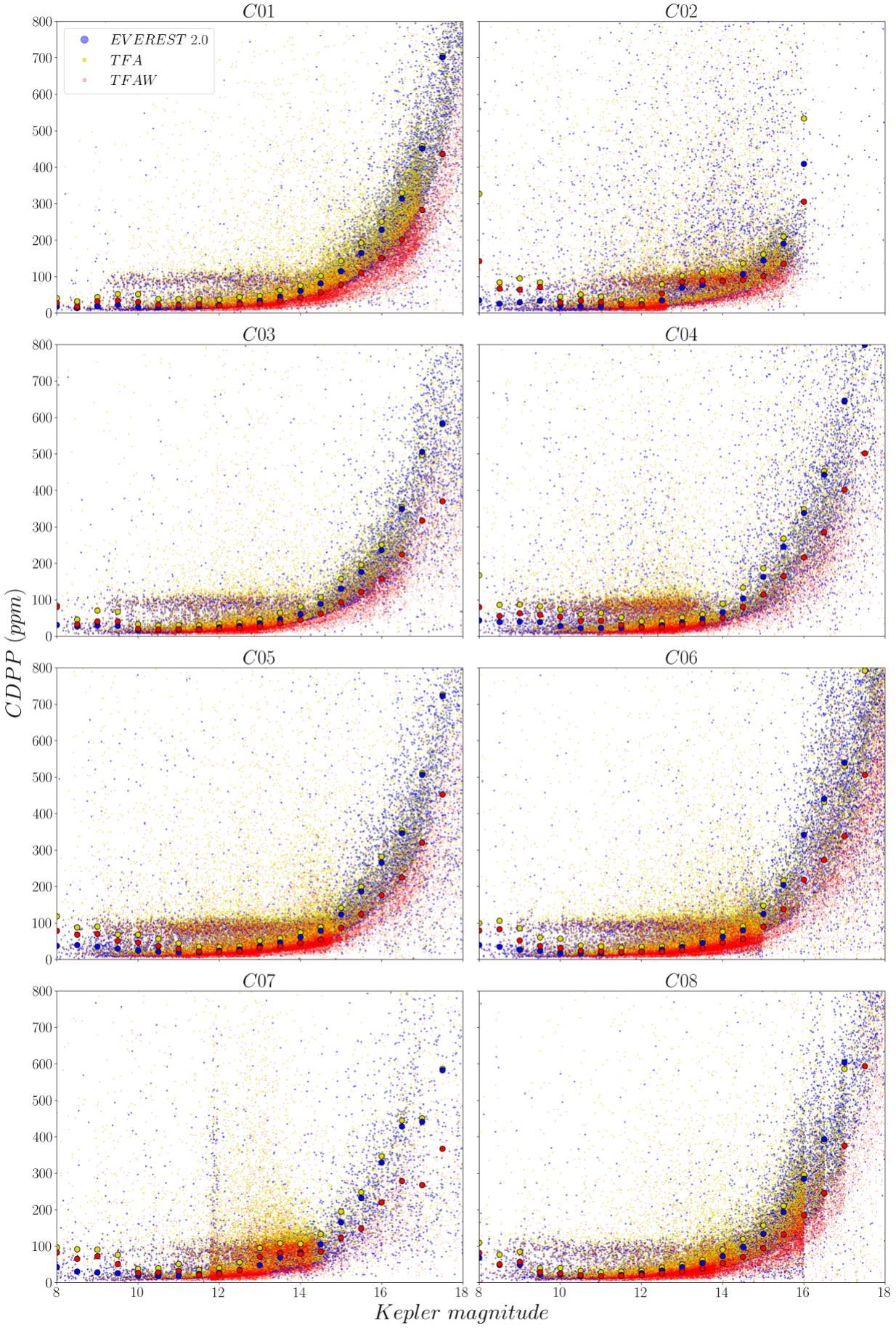}
\caption{6 hr CDPP as a function of $K_p$ for all K2 targets in campaigns C1-C8 corrected with \texttt{EVEREST 2.0} (blue), \texttt{TFA} (yellow) and \texttt{TFAW} (red). The median in 0.5 magnitude-wide bins is indicated by blue circles for \texttt{EVEREST 2.0}, by yellow circles for \texttt{TFA} and red circles for \texttt{TFAW}.}
\label{fig:cdpp_vs_kp_1-8_all}
\end{figure*}

\begin{figure*}
\centering
\includegraphics[height=18cm,keepaspectratio]{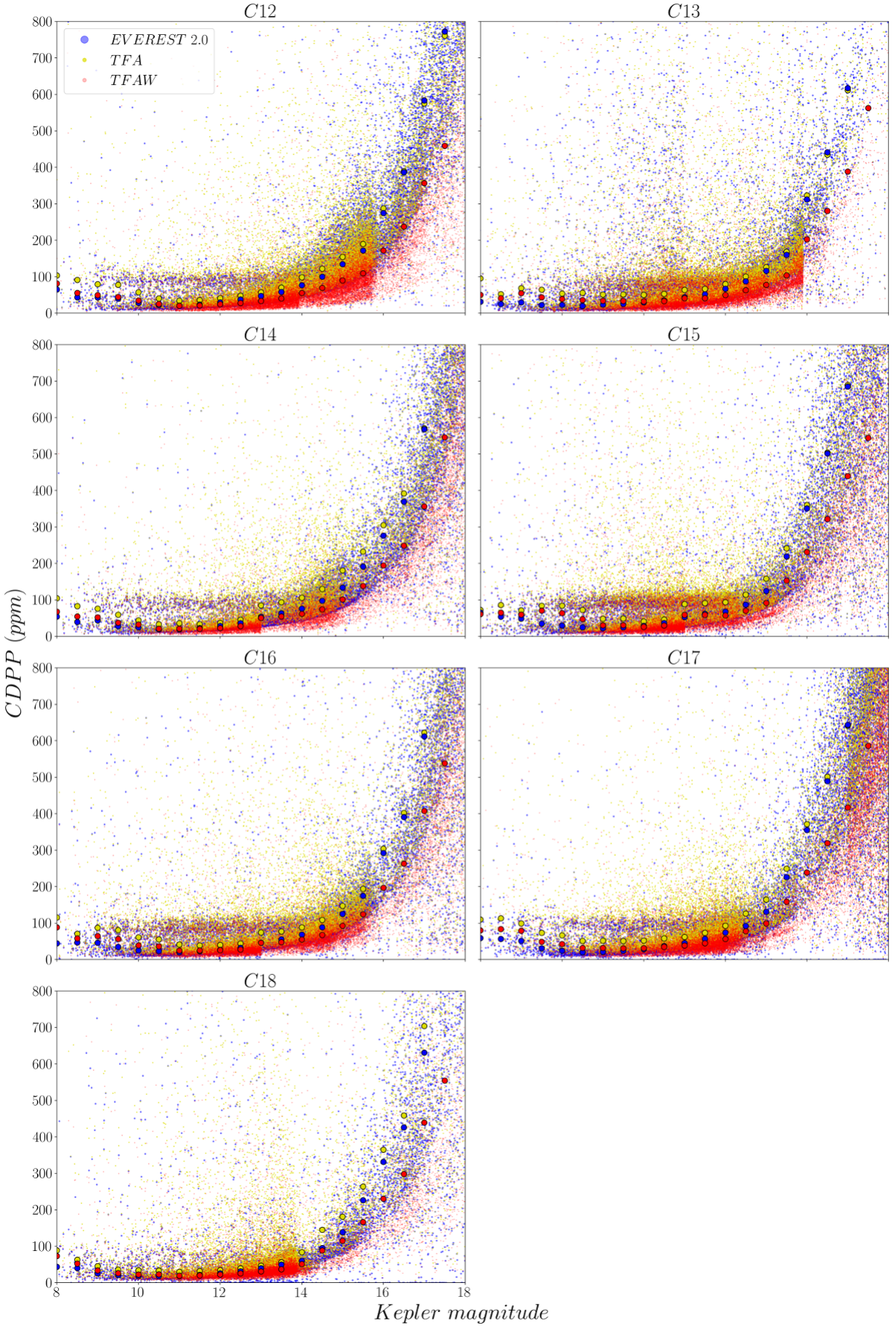}
\caption{6 hr CDPP as a function of $K_p$ for all K2 targets in campaigns C12-C18 corrected with \texttt{EVEREST 2.0} (blue), \texttt{TFA} (yellow) and \texttt{TFAW} (red). The median in 0.5 magnitude-wide bins is indicated by blue circles for \texttt{EVEREST 2.0}, by yellow circles for \texttt{TFA} and red circles for \texttt{TFAW}.}
\label{fig:cdpp_vs_kp_12-18_all}
\end{figure*}

The 6 hr CDPP values computed for all \texttt{EVEREST 2.0} (blue dots), \texttt{TFA} (yellow dots) and \texttt{TFAW} (red dots) K2-detrended light curves are shown in Figures~\ref{fig:cdpp_vs_kp_1-8_all} (campaigns C1-C8) and~\ref{fig:cdpp_vs_kp_12-18_all} (campaigns C12-C18). 

In accordance with the results obtained in \citet{delSer2018}, the 6 hr CDPPs of \texttt{TFA} light curves clearly underperform compared to the \texttt{TFAW} ones. They are also worse when compared to the CDPPs obtained with \texttt{EVEREST 2.0} for most $K_p$ magnitudes and most campaigns. In the best scenarios, \texttt{TFA} is only able to give similar CDPPs to the ones obtained by \texttt{EVEREST 2.0}. Given that \texttt{TFAW} outperforms \texttt{TFA} in all campaigns and magnitude ranges, we focus on the former for other performance comparisons in this paper.

\begin{figure*}
\centering
\includegraphics[height=18cm,keepaspectratio]{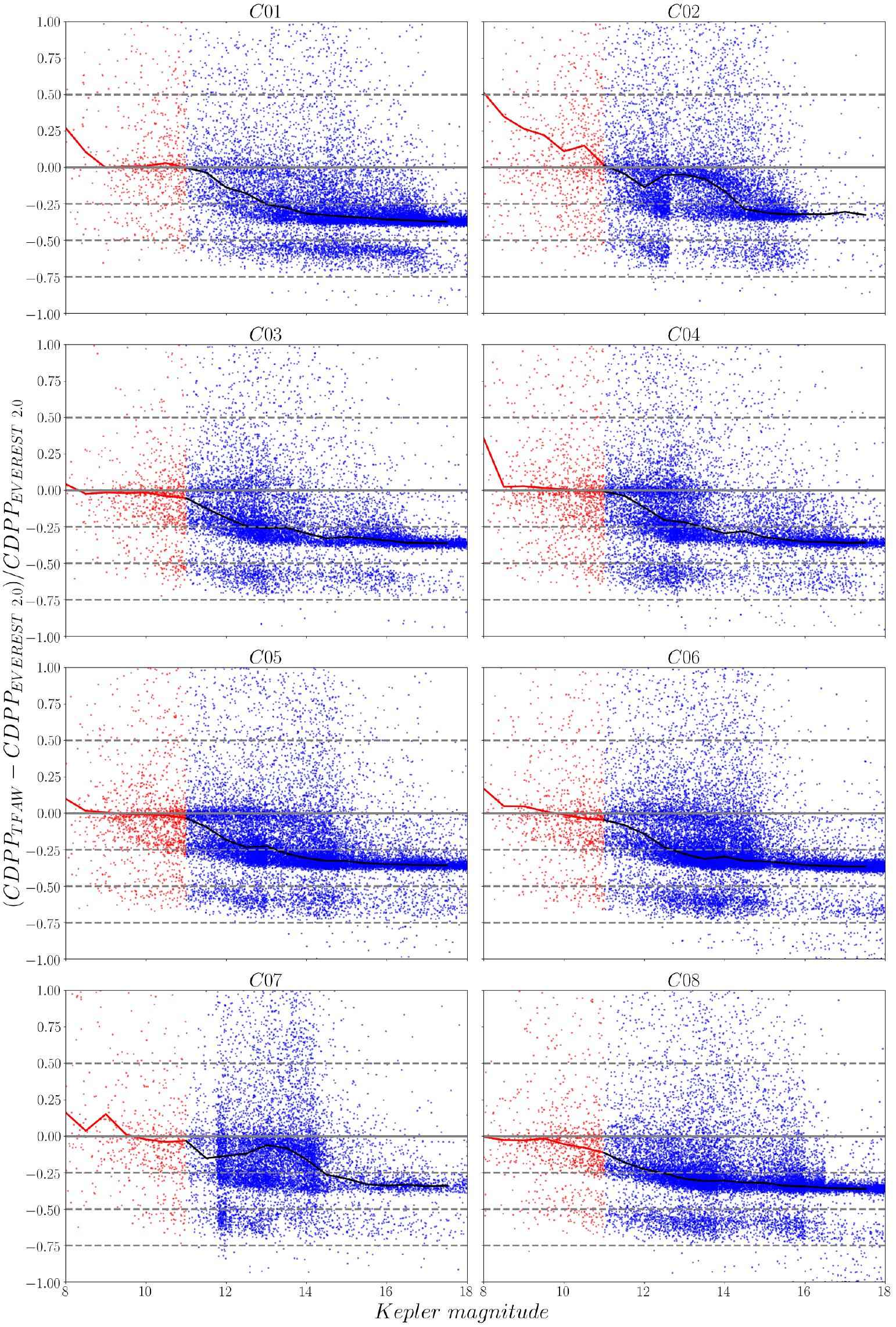}
\caption{\texttt{TFAW} vs. \texttt{EVEREST 2.0} relative 6 hr CDPP comparison for all K2 stars in campaigns C1 to C8. Individual stars are plotted as points, red for saturated stars ($K_p\lesssim$11 mag) and blue for fainter. Median relative 6 hr CDPP is plotted by a solid red line, for saturated stars ($K_p\lesssim$11 mag) and a solid black line for fainter magnitudes.}
\label{fig:ratio_cdpp_vs_kp_1-8_all}
\end{figure*}

\begin{figure*}
\centering
\includegraphics[height=18cm,keepaspectratio]{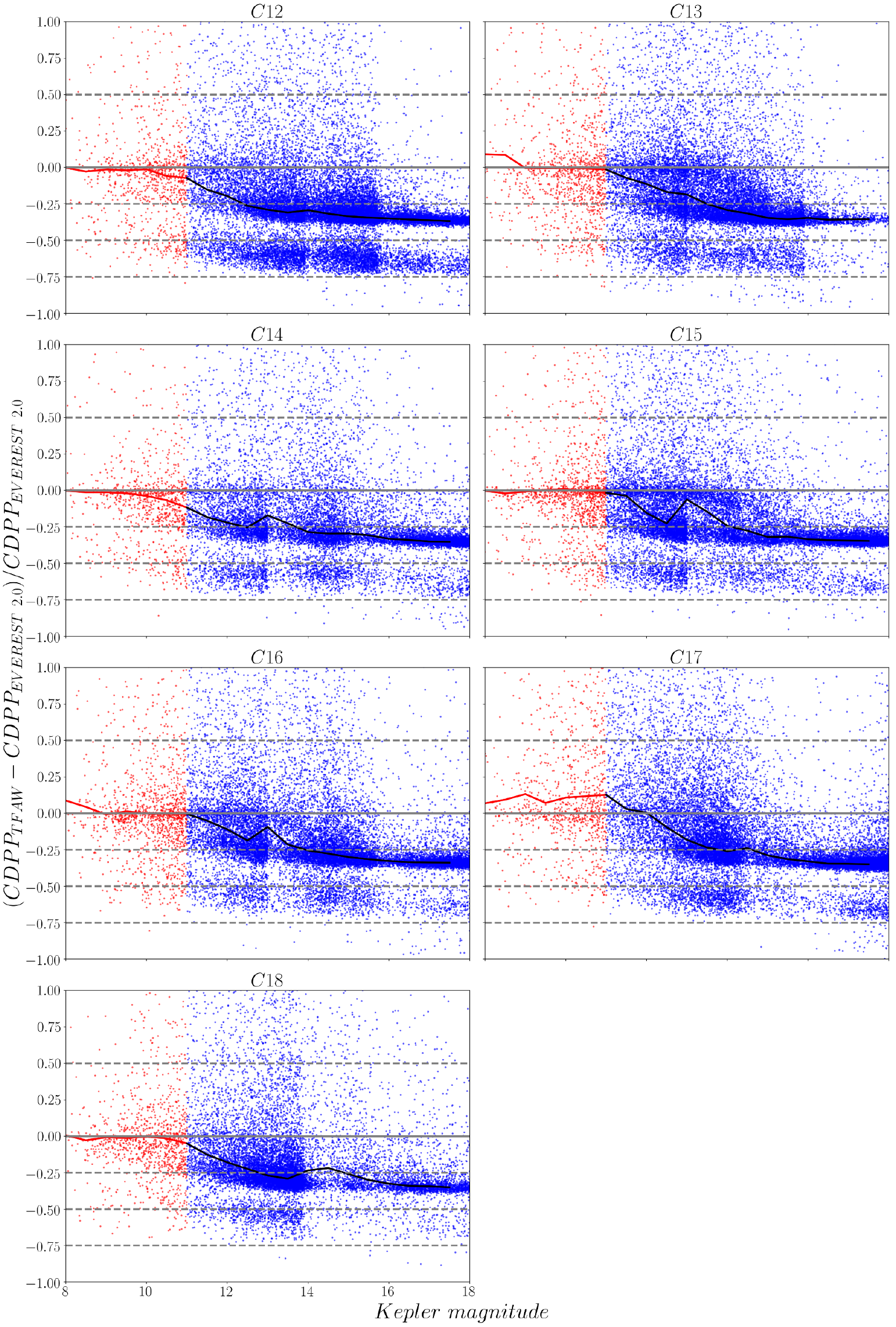}
\caption{\texttt{TFAW} vs. \texttt{EVEREST 2.0} relative 6 hr CDPP comparison for all K2 stars in campaigns C12 to C18. Individual stars are plotted as points, red for saturated stars ($K_p\lesssim$11 mag) and blue for fainter. Median relative 6 hr CDPP is plotted by a solid red line, for saturated stars ($K_p\lesssim$11 mag) and a solid black line for fainter magnitudes.}
\label{fig:ratio_cdpp_vs_kp_12-18_all}
\end{figure*}

Figures~\ref{fig:ratio_cdpp_vs_kp_1-8_all} (campaigns C1-C8) and~\ref{fig:ratio_cdpp_vs_kp_12-18_all} (campaigns C12-C18) show the relative 6 hr CDPP differences between the \texttt{TFAW} corrected light curves and those produced by \texttt{EVEREST 2.0} as a function of $K_p$ magnitude. Individual CDPP values for each star are plotted as blue points and the median in 0.5 magnitude-wide bins as a black solid line. Saturated stars ($K_p\lesssim$11, \citet{Luger2018}) are plotted as red points, with their median indicated by a red solid line. Light curves which have undergone \texttt{TFAW} iterative reconstruction step (i.e. those with a periodicity with SDE$\geq$9.0) and those which have only had removed a first SWT estimation of the high frequency noise have been included in the plots.

On average, the \texttt{TFAW} median relative 6 hr CDPP is $\sim$30\% better than the one from \texttt{EVEREST 2.0} for stars within $11<K_p<15$. For saturated stars (i.e $K_p<$11), plotted as red dots, \texttt{TFAW} yields similar results as \texttt{EVEREST 2.0}, though many of the stars benefit from a slight improvement of about $\sim$5-10\% better CDPP values. For bright stars with 11$<K_p\lesssim$12.5 \texttt{TFAW} light curves have higher precision than those of \texttt{EVEREST 2.0} by $\sim$5-25\%. This improvement increases as we go towards fainter magnitudes and can reach about $\sim$35-40\% better precision. This average \texttt{TFAW} performance is slightly worse for two campaigns: C2 for the larger fraction of (variable) giant stars, and C7 due to a change in the orientation of the spacecraft and excess of jitter. Overall, \texttt{TFAW} improves the photometric precision of \texttt{EVEREST 2.0} light curves for all campaigns and all $K_p$ magnitudes, showing the robustness of \texttt{TFAW} to denoise light curves with different noise properties and coming from different stellar populations.

We also note in Figures~\ref{fig:ratio_cdpp_vs_kp_1-8_all} and~\ref{fig:ratio_cdpp_vs_kp_12-18_all} that the relative 6 hr CDPP appears to be decoupled in three horizontal bands: one following the median, another close to zero relative CDPP and the last one below the median with less points but evenly distributed for all $K_p$ magnitudes. All light curves pertaining to this lower CDPP population have SDE$_{\rm TLS}\geq$9.0 and thus, they have undergone the iterative signal reconstruction and denoising. For these objects, \texttt{TFAW} has removed most of the high frequency noise contribution leading to a $\sim$50-75\% improvement in their CDPPs. The population following the median, is comprised by those light curves with SDE$_{\rm TLS}<$9.0, that have only a first SWT estimation of the noise removed from them during \texttt{TFAW}'s signal detection step (see Section~\ref{sec:tfaw}). Finally, the population close to zero relative CDPP corresponds to the horizontal branch also observed in the CDPP vs $K_p$ plots in Figures~\ref{fig:cdpp_vs_kp_1-8_all} and~\ref{fig:cdpp_vs_kp_12-18_all}. This clump of stars are giants~\citep{Christiansen2012} with short-timescale pulsations that have not been filtered by \texttt{TFAW}'s SWT high-frequency noise estimation. This pulsations are not efficiently captured by the high-pass filter applied during the CDPP computation \citep{Luger2016} leading to higher values. It has to be mentioned that we have run \texttt{TFAW} denoising using only the lowest SWT decomposition level (see Section~\ref{sec:tfaw}) to minimize the possibility of removing high frequency signals of stellar origin that could be of interest. However, the use of more SWT levels could benefit the CDPPs of this giant population (and of all the other stars in general) as they would remove noise/signals within a broader frequency range.

\subsection{Transit detection efficiency}
\label{subsect:transitefficiency}

To assess the transit recovery rate we generate two sets of 5,000 randomly distributed test light curves covering magnitudes from $K_p$=8 to $K_p$=18, one simulating \texttt{EVEREST 2.0} light curves and, the other, simulating light curves during  \texttt{TFAW}'s frequency analysis step. The noise contribution for each of these light curves is estimated by fitting the \texttt{EVEREST 2.0} and \texttt{TFAW} 6 hr CDDPs from Figures~\ref{fig:cdpp_vs_kp_1-8_all} and~\ref{fig:cdpp_vs_kp_12-18_all}. After random white noise is injected, for each light curve at a given $K_p$ value, its corresponding CDPP value is randomly distributed, within a given width, around the fitted value.

To each of these light curves, we inject a transit signal selected from a random distribution of Sun-like (0.7-1.35$R_{\odot}$ and 0.8-1.5$M_{\odot}$) and Earth-like (0.5-2.3$R_{\oplus}$) systems with a random distribution of periods and transit epochs (ensuring at least two detectable transits), solar-like quadratic limb darkening coefficients, orbit inclinations and eccentricities.

We want to compare the transit detection rates obtained with \texttt{EVEREST 2.0} and \texttt{TFAW} light curves using \texttt{TLS}. We use the following criteria to define a significant detection: firstly, the highest peak in the \texttt{TLS} power spectrum must have a period between $[\mathrm{P}_\mathrm{s}-0.01, \mathrm{P}_\mathrm{s}+0.01]$, where $\mathrm{P}_\mathrm{s}$ is the period of the simulated planetary transit, and secondly, the SDE$_{\rm TLS}$ of the highest peak in the power spectrum must be greater than 9.

We run \texttt{TLS} using the stellar limb darkening coefficients associated to each of the light curves and the transit search is done in the (0.01, 31.5) days range. We count the number of times the signal is detected in the \texttt{TFAW} light curves but not detected in the \texttt{EVEREST 2.0} ones and the opposite test. Table~\ref{tab:tfawdetect} shows the number of non-simultaneous detections (both in absolute and percentage) for the 5,000 simulated Earth-Sun-like systems along the $K_p$ range for both \texttt{EVEREST 2.0} and \texttt{TFAW} light curves. The number of non-simultaneous detections for the case of \texttt{TFAW} is a factor $\sim$8.5$\times$ higher than for the case of \texttt{EVEREST 2.0} light curves. Also, the mean SDE$_{\rm TLS}$ values are higher for \texttt{TFAW} than for \texttt{EVEREST 2.0}. We also checked the false probability detection rate by seeing how many times the highest, non-aliased to the injected period peak in the power spectra crossed the threshold. We find similar results for \texttt{EVEREST 2.0} and \texttt{TFAW}: 72 and 80 false detections, respectively.

\begin{table}
\centering
\caption{Mutually exclusive detections and mean SDE values for 5,000 simulated transits in Earth-Sun-like systems. $\mathrm{N}_\mathrm{EVEREST}$: not detected using \texttt{TFAW} light curves, but detected using \texttt{EVEREST 2.0} data. $\mathrm{N}_\mathrm{TFAW}$: detected using \texttt{TFAW} light curves, but not detected using \texttt{EVEREST 2.0} data. $\mathrm{N}_\mathrm{mut}$: simultaneous detections with \texttt{EVEREST 2.0} and \texttt{TFAW}. $\mathrm{SDE}_\mathrm{EVEREST}$: mean \texttt{EVEREST 2.0} SDE$_{\rm TLS}$. $\mathrm{SDE}_\mathrm{TFAW}$: mean \texttt{TFAW} SDE$_{\rm TLS}$. Percentage values in parenthesis are with respect to the 5,000 tested transits.}
\label{tab:tfawdetect}      
\begin{tabular}{c c c c c}       
\hline
\noalign{\smallskip}
$\mathrm{N}_\mathrm{EVEREST}$ & $\mathrm{N}_\mathrm{TFAW}$ & $\mathrm{N}_\mathrm{mut}$ & $\mathrm{SDE}_\mathrm{EVEREST}$ & $\mathrm{SDE}_\mathrm{TFAW}$\\     
\hline                       
\noalign{\smallskip}
80 & 681 & 2652 & 22.48 & 24.71\\
(1.6$\%$) & (13.6$\%$) & (53.1$\%$) & & \\
\hline                                   
\end{tabular}
\end{table}

Table~\ref{tab:tfawranges} shows the distribution of the \texttt{EVEREST 2.0} and \texttt{TFAW} detections in three $K_p$ magnitude bins. In the bright-end regime \texttt{TFAW} shows 2.8$\times$ the number of detections of \texttt{EVEREST 2.0}. In the mid-$K_p$ range (11.0 $< K_p <$ 15.0), \texttt{TFAW} performs 8.9$\times$ better than \texttt{EVEREST 2.0}. Finally, in the faint-end case ($K_p >$15.0), \texttt{TFAW} detects the transit in 21.1$\times$ more light curves than \texttt{EVEREST 2.0}. This increase in the detection rate is in accordance with the improvement in the CDPP values for mid- and faint-magnitude range obtained with \texttt{TFAW} (see Figures~\ref{fig:ratio_cdpp_vs_kp_1-8_all} and \ref{fig:ratio_cdpp_vs_kp_12-18_all}).

\begin{table*}
\caption{Detection distributions of ($\mathrm{N}_\mathrm{EVEREST}$, $\mathrm{N}_\mathrm{TFAW}$, $\mathrm{N}_\mathrm{mut}$) for the 5,000 simulated Earth-Sun-like systems as per three bins of $K_p$ magnitude.}
\label{tab:tfawranges}
\centering
\begin{tabular}{cllcllcll}
\hline
\noalign{\smallskip}
\multicolumn{3}{c}{$K_p <11.0$} & \multicolumn{3}{c}{$11.0< K_p <15.0$} & \multicolumn{3}{c}{$K_p >15.0$} \\ \hline
\multicolumn{1}{l}{$\mathrm{N}_\mathrm{EVEREST 2.0}$} & $\mathrm{N}_\mathrm{TFAW}$ & \multicolumn{1}{l|}{$\mathrm{N}_\mathrm{mut}$} & \multicolumn{1}{l}{$\mathrm{N}_\mathrm{EVEREST 2.0}$} & $\mathrm{N}_\mathrm{TFAW}$ & \multicolumn{1}{l|}{$\mathrm{N}_\mathrm{mut}$} & \multicolumn{1}{l}{$\mathrm{N}_\mathrm{EVEREST 2.0}$} & $\mathrm{N}_\mathrm{TFAW}$ & $\mathrm{N}_\mathrm{mut}$ \\ \hline
32 & \multicolumn{1}{c}{90} & \multicolumn{1}{c|}{1290} & 36 & \multicolumn{1}{c}{338} & \multicolumn{1}{c|}{1283} & 12 & \multicolumn{1}{c}{253} & \multicolumn{1}{c}{79}
\\ \hline
\end{tabular}
\end{table*}

\subsection{Transit injections}

\begin{figure}
\centering
\includegraphics[width=\linewidth,keepaspectratio]{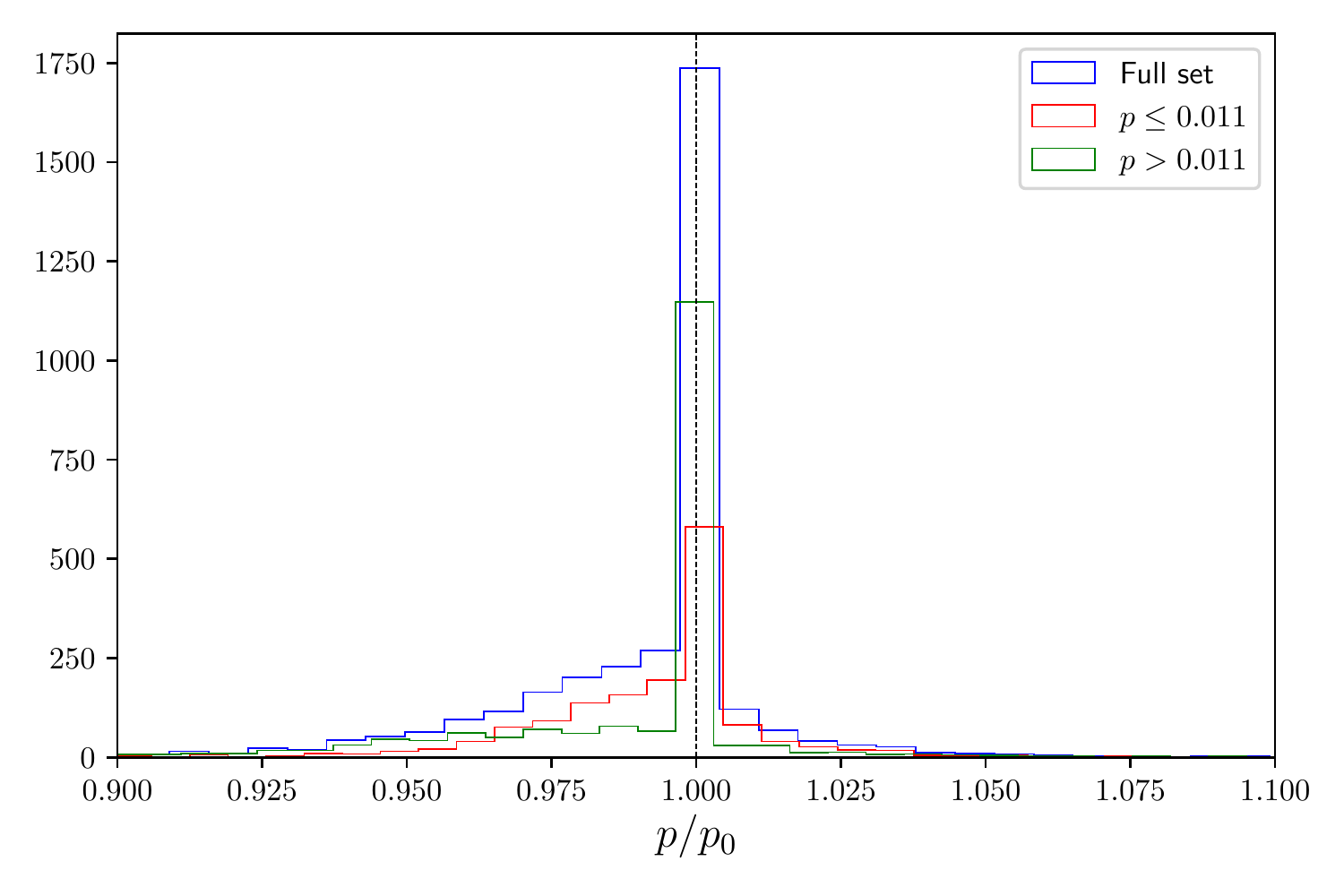}
\caption{Histogram of transits with a certain planetary radius to stellar radius ratio as a fraction of the injected one ($p/p_{0}$) recovered after \texttt{TFAW}'s signal reconstruction step. Blue histogram corresponds to the 3,400 recovered transits; red histogram corresponds to those injected transits with $p\le$0.011 and the green histogram corresponds to transits with $p$>0.011.}
\label{fig:depth_bias}
\end{figure}

In \citet{delSer2018} we show that, thanks to the wavelet approximation of the signal, \texttt{TFAW} is able to diminish the bias in the transit parameters. In order to ensure that \texttt{TFAW} returns an unbiased set of transit parameters, we run a transit injection/recovery test similar to the one used by \citet{Luger2016}. We use a set of 2-day Savitsky-Golay filtered \texttt{EVEREST 2.0} real K2 light curves with no known transit. We randomly select a sample of 3,400 stars from campaigns C1-C8 and C12-C18 with 8$\le$$K_{p}$$\le$18 magnitudes. Using the \texttt{batman} \footnote{\url{https://www.cfa.harvard.edu/~lkreidberg/batman/}} package \citep{Kreidberg2015}, the cataloged stellar properties (i.e. stellar mass and radius and quadratic limb darkening coefficients) for each target, assuming circular orbits, and randomly selecting the transit parameters (planetary radius to stellar radius ratio, orbital period, orbit inclination and transit epoch), we inject one planet (from a hot Jupiter to an Earth-sized planet) transit into each selected \texttt{EVEREST 2.0} light curve. The injected transit depths range from $\sim5\cdot10^{-5}$ to $\sim10^{-2}$. We then run \texttt{TFAW} to reconstruct and denoise the light curves. 
To determine whether the \texttt{TFAW}-corrected light curves can bias the transit depths, we fix all the parameters except for the planetary radius to stellar radius ratio, $p$, at their true values and recover the later. We fit the transit depth of the \texttt{TFAW}-corrected light curves by minimizing the residuals using the Levenberg-Marquardt method implemented in the \texttt{lmfit} package\footnote{\url{https://doi.org/10.5281/zenodo.3814709}}.
Figure \ref{fig:depth_bias} shows the histogram of the recovered planetary radius to stellar radius ratio as a fraction of the injected one ($p/p_{0}$). As can be seen, the median $p/p_{0}$ for \texttt{TFAW}-reconstructed transits, is consistent with 1.0. We find an small <$\sim$2.5$\%$ bias towards smaller ratios for some of the transits. This bias starts to be significant for $p$<$\sim$0.011 (i.e transit depth smaller than $\sim10^{-4}$). However, the relative difference between $p_{0}$ and the recovered planetary radius to stellar radius ratio is smaller than $\sim$0.0005 for almost all ($\sim$97$\%$) the simulated transits.

\subsection{Characterization of known planets}
\label{subsect:characknown}

In \citet{delSer2018} we show that \texttt{TFAW} can improve the MCMC posterior distributions, diminish the bias in the fitted transit parameters and their uncertainties and narrow the credibility intervals for simulated transits. In this Section we compare \texttt{EVEREST 2.0} and \texttt{TFAW} performance in terms of assessing the bias of the MCMC fitted transit parameters values and their uncertainties for two confirmed planetary systems with different SNRs: K2-44 b and K2-298 b.

\subsubsection{Data description}

As the starting point for the computation of \texttt{TFAW}, we use the PLD, CBV-corrected fluxes provided by the \texttt{EVEREST 2.0} pipeline for both targets and all stars present in the same campaign and CCD module. 3072 epochs with the \texttt{QUALITY}=0 flag are considered and a 2-day Savitsky-Golay filter and a 5$\sigma$ outlier clipping are applied to the light curves. To iteratively denoise and reconstruct the light curves with \texttt{TFAW}, a template of reference stars for each target ($\sim$90) is built from a sub-set of stars from the same CCD module using Stetson's L variability index \citep{Stetson1996} to avoid the inclusion of bona-fide variable stars in the sample. We use this \texttt{TFAW}-corrected light curve to run the MCMC fit and compare it to the one obtained from the \texttt{EVEREST 2.0} light curve. 

Prior to the \texttt{TFAW} analysis, for both K2-44 b and K2-298 b we checked that the light curves we obtained directly from the \texttt{EVEREST 2.0} matched with the ones after the transits have been masked. This way we ensure that any bias in the depth of the transiting planet is minimized \citep{Luger2018}.

\subsubsection{Transit parameters fitting procedure}
\label{subsubsect:k2MCMC}

To characterize the target transits we use the analytic transiting model provided by the \texttt{batman} package with quadratic limb darkening coefficients as per \citet{Mandel2002}. We assume circular orbits (i.e. eccentricity=0) and fit the following five transit parameters: the transit epoch, $T_0$, the orbital period, $P$, the semi-major axis of the orbit, $a$, the planetary radius to stellar radius ratio, $p$, and the inclination of the orbit, $i$. We use the MCMC sampler provided by the \texttt{emcee} \citep{Foreman2013} package and use the \texttt{george} \citep{Ambikasaran2015} package to create a combined model consisting on a Matérn 3/2 kernel plus a jitter or "white" noise term to generalize the likelihood function in order to consider covariances between data points (i.e correlated noise) and to minimize the bias of the inferred parameters. We consider a uniform distribution of the priors with wide enough bounds to let the chains explore the parameter space without getting close to the bound limit: $\pm1$ day around the cataloged transit epoch for $T_0$, $\pm3$ days for $P$, from 2$R_{*}$ to the cataloged semi-major axis plus 5$R_{*}$ for $a$, the cataloged planet/star ratio $\pm0.01$ for $p$, and from 85$^{\circ}$ to 90$^{\circ}$ for $i$. We run the sampler with 100 walkers, 10,000 iterations with a burn-in phase of 2,000 iterations. This way we ensure that each of the chains run for more than 50 auto-correlation times for each parameter and that the mean acceptance fraction is between 0.25 and 0.5 \citep{Bernardo1996,Foreman2013}.

\subsubsection{K2-44: a confirmed high-SNR single planetary system example}
\label{subsubsec:k2-44}

K2-44 (EPIC 201295312) is a V=12.19$\pm$0.12 mag \citep{Zacharias2012} located at $(\alpha,~\delta)$ = (11:36:02.79, -02:31:15.17) \citep{Gaia2018}. It was observed by the K2 mission during the C1 monitoring campaign from May 30 to Aug 21, 2014. K2-44 was first reported as a planetary hosting candidate by \citet{Montet2015}, and later validated, confirmed and characterized by \citet{Crossfield2016} and \citet{Mayo2018}. For a more detailed summary of the stellar and planetary parameters see Table \ref{tab:K2-44b}.

\begin{table}
\centering
\caption{Stellar and planetary parameters obtained for K2-44 b by \citet{Crossfield2016}, \citet{Sing2010}, \citet{Mayo2018}.} 
\label{tab:K2-44b}
\begin{tabular}{ll}
\hline
K2 ID & EPIC 201295312 \\ 
\hline
Stellar parameters &  \\
Stellar radius $R_{\rm s}$ ($R_{\odot}$) & $1.58\pm0.15$ \\
Stellar mass $M_{\rm s}$ ($M_{\odot}$) & $1.150\pm0.060$ \\
Effective temperature (K) & $5912\pm51$ \\
Surface gravity ($log_{10}(cm/s^2)$) & $4.101\pm0.063$ \\
Metallicity [Fe/H] & $0.0$ (assumed) \\
Spectral Type & - \\ \hline
Transit parameters &  \\
Period $P$ (days) & $5.65688\pm0.00059$ \\
Transit epoch $T_0$ (BJD - 2454833) (days) & $1978.7176\pm0.0044$ \\
Transit duration (hours) & $4.36\pm0.13$ \\
Eccentricity $e$ & 0 (assumed) \\
Radius ratio $p$ & $0.0156\pm0.0012$ \\
$q_{1}$ & $0.4752^{\dagger}$ \\
$q_{2}$ & $0.1914^{\dagger}$ \\
Scaled semi-major axis $a$ (AU) & $0.0651\pm0.0011$ \\
Inclination $i$ ($^{\circ}$) & $87.354350^{+1.856108}_{-3.300347}$ $^{\star}$ \\
\hline
Planetary parameters &  \\
Planetary radius $R_{p}$ ($R_{\oplus}$) & $2.72\pm0.32$ \\
\hline
\multicolumn{2}{l}{$^{\dagger}$ denote values from \citet{Claret2018} assuming 0.0 [Fe/H] metallicity.}\\
\multicolumn{2}{l}{$^{\star}$ denote values from \citet{Mayo2018}.}\\
\end{tabular}
\end{table}

To check whether the improved photometric precision yielded by \texttt{TFAW} in Section~\ref{subsect:cdpp_allcvs} results in a better characterization of the transiting signal we analyze \texttt{EVEREST 2.0}- and \texttt{TFAW}-corrected light curves for the confirmed exoplanet K2-44 b and compare the fitted parameters with the ones obtained by \citet{Crossfield2016} and \citet{Mayo2018}. We assume a circular orbit ($e=0$), and a longitude of the periastron of $\omega$=90$^{\circ}$. Using the stellar parameters provided by \citet{Crossfield2016} (see Table \ref{tab:K2-44b}) and, assuming a metallicity of [Fe/H]=0.0, we fix the quadratic limb darkening coefficients to their theoretical values taken from \citet{Claret2018}.

In Table \ref{tab:K2-44b_comp} we compare the transit parameters and their uncertainties obtained by \citet{Crossfield2016} and \citet{Mayo2018} with the ones obtained with \texttt{EVEREST 2.0} and \texttt{TFAW} posterior probability distributions after running the MCMC fit as indicated in Section~\ref{subsubsect:k2MCMC}. The fitted parameter values are obtained from the 50\% quantiles and their upper and lower errors are computed from the 25\% and 75\% quantiles, respectively.

   \begin{table*}
   \caption{Top table: K2-44 b parameters from \citet{Crossfield2016},~\citet{Mayo2018}, and posterior transit parameters values and their uncertainties (25\% and 75\% quantiles) for \texttt{EVEREST 2.0} and \texttt{TFAW} MCMC fits. Middle table: 95\% confidence intervals of the highest probability density for K2-44 b transit parameters \texttt{EVEREST 2.0} and \texttt{TFAW} MCMC fits. Bottom table: Derived parameters from \citet{Crossfield2016},~\citet{Mayo2018}, \texttt{EVEREST 2.0}, and \texttt{TFAW}.} 
   \label{tab:K2-44b_comp}
   \centering                         
   \begin{tabular}{c c c c c c}       
   \hline
   \noalign{\smallskip}
MCMC Parameters                              & $T_0$ (BJD-2454833)                & $P$ (days)            & $a$ ($AU$)                     & $p$                               & $i$ ($^{\circ}$) \\ \hline
\noalign{\smallskip}
   \citet{Crossfield2016} & 1978.7176$\pm$0.0044 & 5.65688$\pm$0.00059 & 0.0651$\pm$0.0011 & 0.0156$\pm$0.0012 & - \\
   \noalign{\smallskip}  
   \citet{Mayo2018} & 1978.72011$^{+0.002565}_{-0.002557}$ & 5.656304$^{+0.000366}_{-0.000323
}$ & - & 0.017257$^{+0.000704}_{-0.000538}$ & 87.354350$^{+1.856108}_{-3.300347}$ \\
   \noalign{\smallskip}  
   \texttt{EVEREST 2.0} & 1978.7248$^{+0.0020}_{-0.0019}$  & 5.6549$\pm$0.0003 &      0.0644$^{+0.0030}_{-0.0048}$ & 0.0149$\pm$0.0005 & 87.7663$^{+1.1671}_{-1.3999}$ \\
   \noalign{\smallskip}  
   \texttt{TFAW} & 1978.7094$^{+0.0012}_{-0.0011}$ & 5.6570$\pm$0.0002 & 0.0600$^{+0.0019}_{-0.0029}$ & 0.0141$^{+0.0005}_{-0.0004}$ & 87.9933$^{+1.0218}_{-1.1633}$ \\
   \noalign{\smallskip}  
   \hline
   \noalign{\smallskip}  
   \multicolumn{6}{c}{95\% confidence intervals of the highest posterior density}\\ 
   \noalign{\smallskip}
   \hline    
   \noalign{\smallskip}
\noalign{\smallskip}  
\texttt{EVEREST 2.0} & 1978.71873 - 1978.73143 & 5.65391 - 5.65598 & 0.05310 - 0.07049 & 0.01342 - 0.01645 & 84.96043 - 89.99988 \\
\noalign{\smallskip}  
\texttt{TFAW} & 1978.70600 - 1978.71302 & 5.65638 - 5.65757 & 0.05308 - 0.06393 & 0.01302 - 0.01556 & 85.66826 - 89.99971 \\
\noalign{\smallskip}  
\hline
\noalign{\smallskip}  
\multicolumn{2}{c}{Derived system parameters} & \multicolumn{2}{c}{$R_{p}$ ($R_{\oplus}$)} & \multicolumn{2}{c}{b} \\ 
\noalign{\smallskip}  
\hline
   \noalign{\smallskip}  
\multicolumn{2}{c}{\citet{Crossfield2016}}  & \multicolumn{2}{c}{2.72$\pm$0.32} & \multicolumn{2}{c}{-} \\
   \noalign{\smallskip}  
\multicolumn{2}{c}{\citet{Mayo2018}} & \multicolumn{2}{c}{$2.93920584211^{+0.496465103254}_{-0.376603037856}$} & \multicolumn{2}{c}{-} \\
   \noalign{\smallskip}  
\multicolumn{2}{c}{\texttt{EVEREST 2.0}} & \multicolumn{2}{c}{$2.571^{+0.089}_{-0.089}$} & \multicolumn{2}{c}{$0.34^{+0.18}_{-0.21}$} \\
\noalign{\smallskip}
\multicolumn{2}{c}{\texttt{TFAW}} & \multicolumn{2}{c}{$2.433^{+0.089}_{-0.072}$} & \multicolumn{2}{c}{$0.29^{+0.15}_{-0.17}$} \\
\noalign{\smallskip}
   \hline                                   
   \end{tabular}
   \end{table*}

\begin{figure*}
\centering
\includegraphics[height=11.0cm,keepaspectratio]{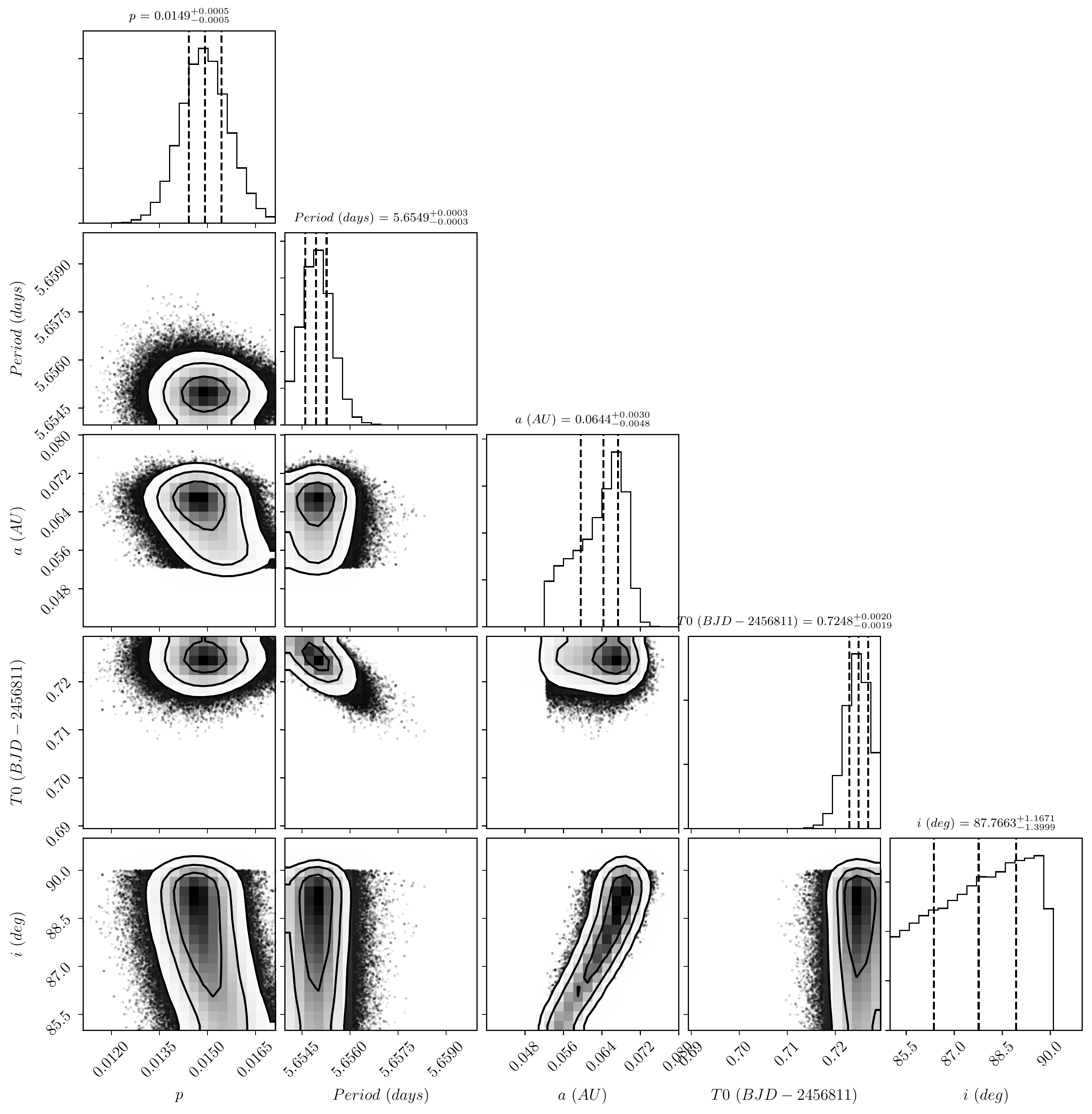}
\includegraphics[height=11.0cm,keepaspectratio]{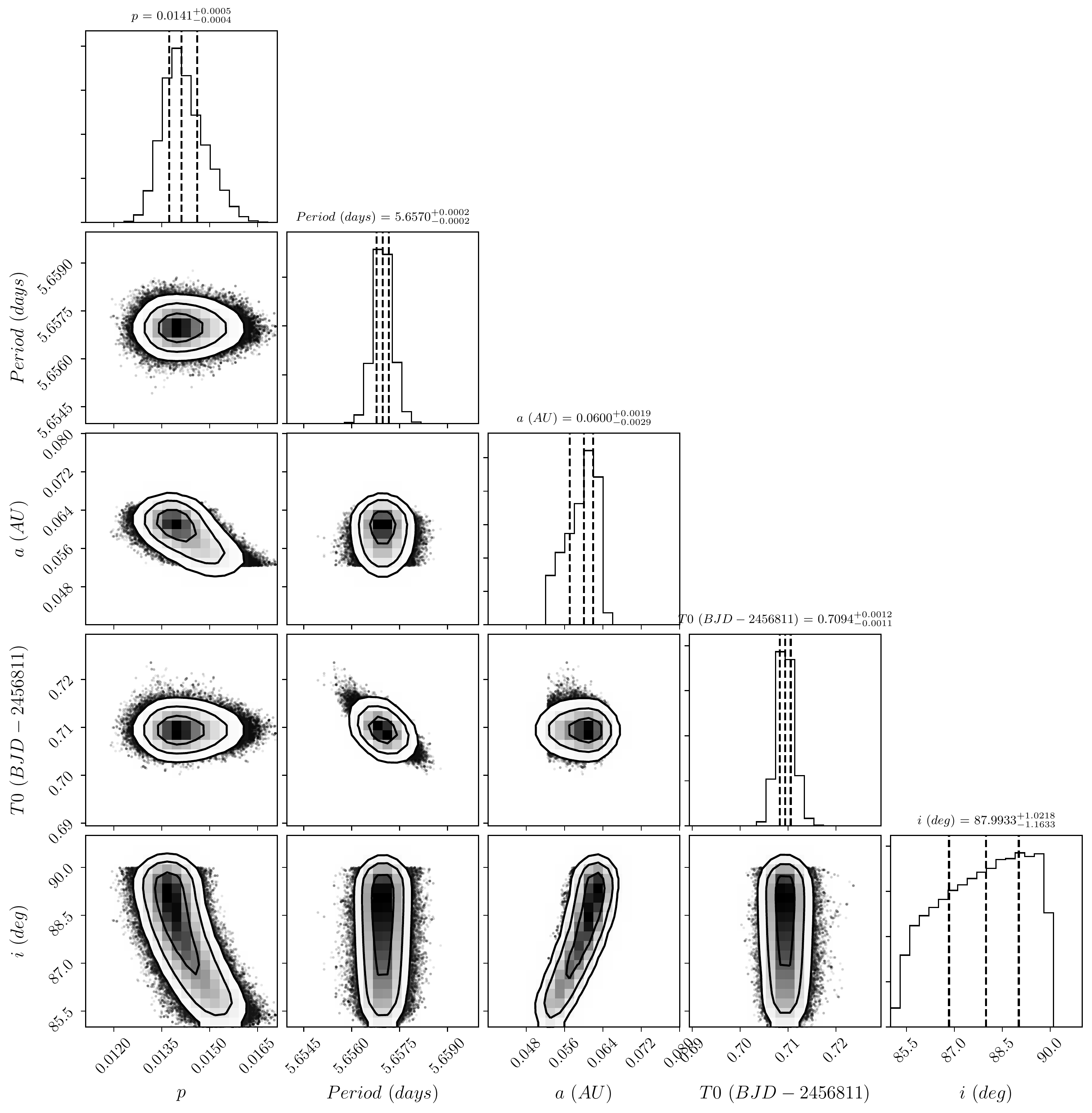}
\caption{1-D and 2-D projections of the posterior probability distributions of the 5 MCMC fitted parameters ($p$, $P$, $a$, $T_{0}$, $i$) for K2-44 b \texttt{EVEREST 2.0} (top) and \texttt{TFAW} (bottom) light curves. The  25\%, 50\%, 75\% quantiles, are displayed in dashed vertical lines on the 1-D histograms, and on the top each panels column.}
\label{fig:K2-44_b_mcmc_corner}
\end{figure*}

The MCMC fit corner plot \citep{Foreman2016} for the K2-44 b transit is shown in Figure~\ref{fig:K2-44_b_mcmc_corner}, yielding the following results:
the time of inferior conjunction, $T_{0}$, for \texttt{EVEREST 2.0} and \texttt{TFAW} are compatible with the one reported by \citet{Crossfield2016} and \citet{Mayo2018}, though for \texttt{TFAW}, the uncertainties are much lower than for the other three ($\sim$2$\times$ for \texttt{EVEREST 2.0}, $\sim$2.5$\times$ for \citet{Mayo2018}, and $\sim$4$\times$ compared to \citet{Crossfield2016}). For the semi-major axis of the orbit, $a$, \texttt{EVEREST 2.0} and \texttt{TFAW} return smaller values than the one by \citet{Crossfield2016}. However, while \texttt{EVEREST 2.0} value is compatible within the errors with the one reported by \citet{Crossfield2016}, this is not the case of \texttt{TFAW}. Though it is still compatible if ones takes into account the uncertainties in the stellar radius, impact parameter and orbit inclination. For the latter, both \texttt{EVEREST 2.0} and \texttt{TFAW} yield values compatible with the reported value by \citet{Mayo2018}. Again, \texttt{TFAW} returns the smallest uncertainties. Regarding the planetary to star radius ratio, $p$, both \texttt{EVEREST 2.0} and \texttt{TFAW} obtain compatible values within the errors with the one reported by \citet{Crossfield2016}. Again, \texttt{TFAW} returns the smallest uncertainties for this parameter and, compatible with the $p/p_{0}$ dispersion seen in Figure \ref{fig:depth_bias}. For the period, $P$, the values from \texttt{TFAW} and \texttt{EVEREST 2.0} are compatible with the cataloged ones. As for the previous parameters, \texttt{TFAW} returns the smallest uncertainties for the period. In summary, even for this rather high-SNR transit, \texttt{TFAW} returns lower uncertainties for all parameters compared with the \texttt{EVEREST 2.0} and cataloged ones. Also, following the results in \citet{delSer2018} with simulated transits, the transit parameters obtained with \texttt{TFAW} might be closer to the real ones (assuming that the planet orbits in a circular orbit). Finally, the widths of \texttt{TFAW}'s 95\% confidence intervals are narrower for all parameters.

With the best fit parameters, \texttt{EVEREST 2.0} obtains a mean planetary radius of $2.571^\pm0.089$ $R_{\oplus}$ and \texttt{TFAW} $2.433^{+0.089}_{-0.072}$ $R_{\oplus}$, both slightly below of the values reported by \citet{Crossfield2016} and \citet{Mayo2018}, but compatible within the errors. Although $b$ is not reported either by \citet{Crossfield2016} or \citet{Mayo2018}, we also derive it from the best fit parameters for \texttt{EVEREST 2.0} to be $0.34^{+0.18}_{-0.21}$, and for \texttt{TFAW} to be $0.29^{+0.15}_{-0.17}$.

\begin{figure*}
\centering
\includegraphics[height=15.0cm,keepaspectratio]{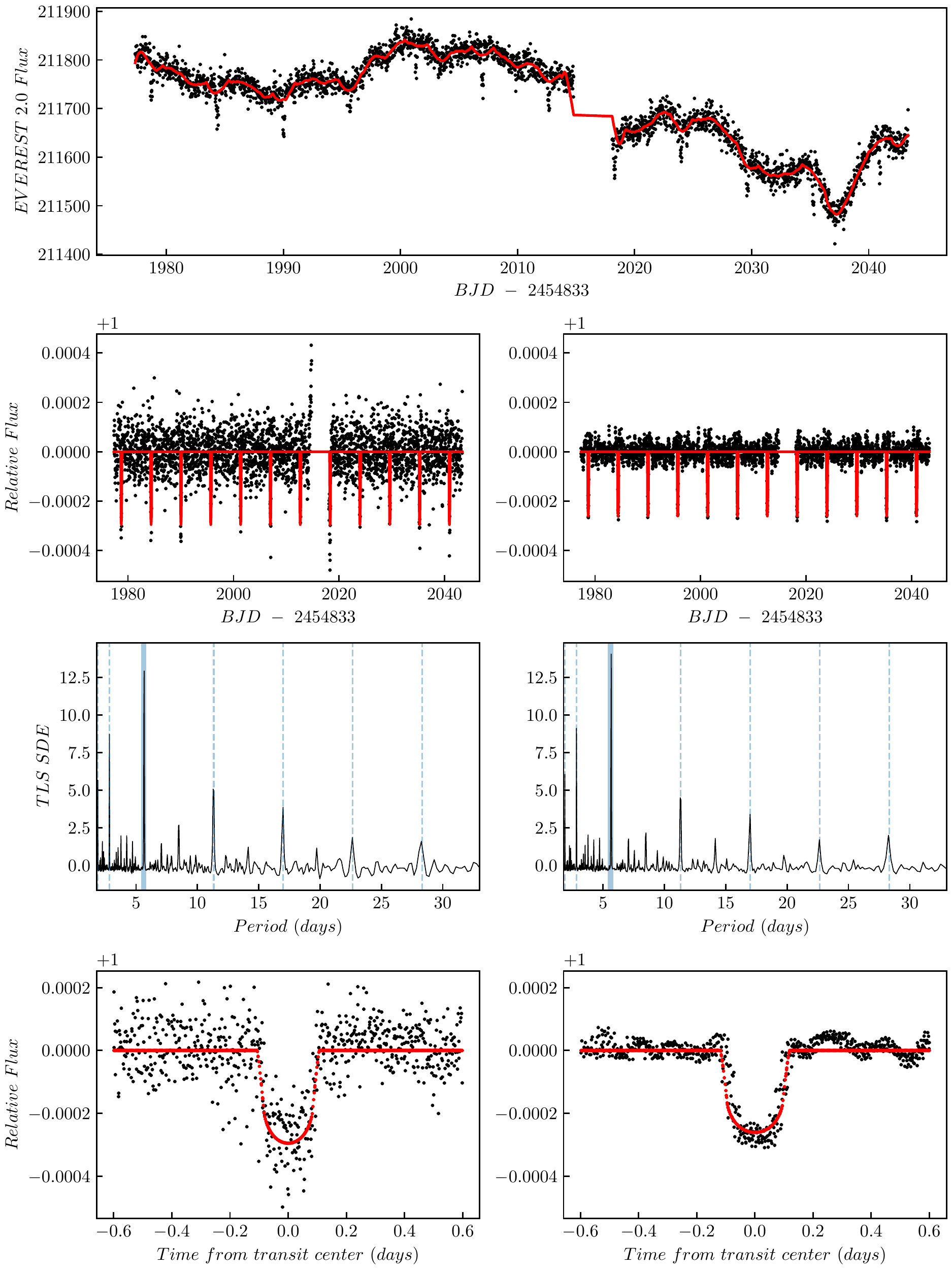}
\caption{\textit{From top to bottom}, the panels show the K2-44 raw flux of the \texttt{EVEREST 2.0} light curve (with the running median in red), the entire light curve with in-transit data of the new candidate marked in red (left for \texttt{EVEREST 2.0} and right for \texttt{TFAW}), the SDE$_{\rm TLS}$ periodogram (left for \texttt{EVEREST 2.0} and right for \texttt{TFAW}), and the normalized phase-folded to the K2-44 b period light curve with MCMC fit data marked in red (left for \texttt{EVEREST 2.0} and right for \texttt{TFAW}).}
\label{fig:k2-44b}
\end{figure*}

Figure~\ref{fig:k2-44b} shows the summary plot displaying the PLD, CBV-corrected flux provided by the \texttt{EVEREST 2.0} pipeline for K2-44 with the 2-day running median plotted in red, the \texttt{EVEREST 2.0} median-filtered light curve and the \texttt{TFAW}-corrected light curves (with the MCMC derived transit data of the new candidate plotted in red), the \texttt{TLS} periodograms for \texttt{EVEREST 2.0} and \texttt{TFAW}'s frequency analysis step; and the phase-folded light curves with the MCMC fit data (red line) for \texttt{EVEREST 2.0} (left) and \texttt{TFAW} iteratively denoised and reconstructed one (right). The \texttt{TLS} periodograms show the position of the confirmed planet detected period (solid blue line) and its harmonics (dashed blue lines).

\subsubsection{K2-298: a confirmed+candidate low-SNR multi planetary system example}
\label{subsec:k2-298}

K2-298 (EPIC201841433) is a V=14.87$\pm$0.01 mag \citep{Zacharias2012} star located at $(\alpha,~\delta)$ = (11:40:49.62, +06:08:05.44) \citep{Gaia2018}. It was observed by the K2 mission during the C1 monitoring campaign from May 30 to Aug 21, 2014. K2-298 was validated as a $0.802^{+0.081}_{-0.163}$ $M_{\odot}$ \citep{Heller2019}, $T_{eff}$=$5053^{+103}_{-166}$ K, $0.62846106^{+0.04341984}_{-0.02485835}$ $R_{\odot}$ \citep{Gaia2018} star orbited by an inner planet, K2-298 b, of $1.10^{+0.14}_{-0.12}$ $R_{\oplus}$ at $11.5^{+5.2}_{-2.6}$ $R_{\rm s}$ \citep{Kruse2019} with a period of $4.16888^{+0.00050}_{-0.00056}$ days \citep{Heller2019}, and an outer candidate, EPIC201841433.01, of $2.11^{+0.23}_{-0.61}$ $R_{\oplus}$  at $29.5^{+14.7}_{-6.7}$ $R_{\rm s}$ with a period of $12.3389^{+0.0016}_{-0.0017}$ days \citep{Kruse2019}. Table \ref{tab:K2-298b} summarizes all the stellar and planetary parameters for K2-298 b. 

As with the K2-44 case, we want to check whether the improved photometric precision yielded by \texttt{TFAW} results in a better characterization of the transiting signal. We analyze \texttt{EVEREST 2.0}- and \texttt{TFAW}-corrected light curves for the confirmed exoplanet K2-298 b and compare the fitted parameters with the ones obtained by \citet{Heller2019} and \citet{Kruse2019}. We, again, assume a circular orbit ($e=0.0$), and a longitude of the periastron of $\omega$=90$^{\circ}$. Using the stellar parameters in Table \ref{tab:K2-298b} and, assuming a metallicity of [Fe/H]=0.0, we fix the quadratic limb darkening coefficients to their theoretical values taken from \citet{Claret2018}.

\begin{table}
\centering
\caption{Stellar and planetary parameters obtained for K2-298 b by \citet{Heller2019}, \citet{Kruse2019}, \citet{Sing2010}, \citet{Gaia2018}, \citet{Andrae2018}.} 
\label{tab:K2-298b}
\begin{tabular}{ll}
\hline
K2 ID & EPIC 201841433 \\ 
\hline
Stellar parameters &  \\
Stellar radius $R_{\rm s}$ ($R_{\odot}$) & $0.62846106^{+0.04341984}_{-0.02485835}$ $^{\star\star}$ \\
Stellar mass $M_{\rm s}$ ($M_{\odot}$) & $0.802^{+0.081}_{-0.163}$ \\
Effective temperature (K) & $5053^{+103}_{-166}$ $^{\star\star}$ \\
Surface gravity ($log_{10}(cm/s^2$)) & $4.595^{+0.050}_{-2.860}$ \\
Metallicity [Fe/H] & $0.0$ (assumed) \\
Spectral Type & - \\ \hline
Transit parameters &  \\
Period $P$ (days) & 4.16959$^{+0.00051}_{-0.00053}$ \\
Transit epoch $T_0$ (BJD - 2454833) (days) & 2020.3300$^{+0.0037}_{-0.0033}$ \\
Transit duration (hours) &  $2.400^{+0.264}_{-0.336}$ $^{\star}$ \\
Eccentricity $e$ & 0 (assumed) \\
Radius ratio $p$ & 0.0160$^{+0.0017}_{-0.0016}$ \\
$q_{1}$ & $0.5510^{\dagger}$ \\
$q_{2}$ & $0.1575^{\dagger}$ \\
Scaled semi-major axis $a$ (AU) & $0.0503^{+0.0227}_{-0.0114}$ $^{\star}$ \\
Inclination $i$ ($^{\circ}$) & $88.21$ $^{\ddagger}$ \\
\hline
Planetary parameters &  \\
Planetary radius $R_{p}$ ($R_{\oplus}$) & $1.10^{+0.14}_{-0.12}$ \\
\hline
\multicolumn{2}{l}{$^{\dagger}$ denotes values from \citet{Claret2018} assuming 0.0 [Fe/H] metallicity.}\\
\multicolumn{2}{l}{$^{\ddagger}$ denotes values derived assuming $b=0.36^{+0.24}_{-0.25}$ \citep{Heller2019}} \\ 
\multicolumn{2}{l}{and $a$=$11.5^{+5.2}_{-2.6}~R_{\rm s}$ \citet{Kruse2019}.}\\
\multicolumn{2}{l}{$^{\star}$ denotes values taken from \citet{Kruse2019}.}\\
\multicolumn{2}{l}{$^{\star\star}$ denotes values derived from \citet{Gaia2018}}\\
\end{tabular}
\end{table}

In Table \ref{tab:K2-298b_comp} we compare the transit parameters and their uncertainties obtained by \citet{Heller2019} and \citet{Kruse2019} with the ones obtained with \texttt{EVEREST 2.0} and \texttt{TFAW} MCMC posterior probability distributions. As with K2-44 b, the fitted parameter values are obtained from the 50\% quantiles and their upper and lower errors are computed from the 25\% and 75\% quantiles, respectively.

   \begin{table*}
   \caption{Top table: K2-298 b parameters from \citet{Heller2019},~\citet{Kruse2019}, and posterior transit parameters values and their uncertainties (25\% and 75\% quantiles) for \texttt{EVEREST 2.0} and \texttt{TFAW} MCMC fits. Middle table: 95\% confidence intervals of the highest probability density for K2-298 b transit parameters \texttt{EVEREST 2.0} and \texttt{TFAW} MCMC fits. Bottom table: Derived parameters from \citet{Heller2019},~\citet{Kruse2019}, \texttt{EVEREST 2.0}, and \texttt{TFAW}.}   
   \label{tab:K2-298b_comp}
   \centering                         
   \begin{tabular}{c c c c c c}       
   \hline
   \noalign{\smallskip}
MCMC Parameters                              & $T_0$ (BJD-2454833)                & $P$ (days)            & $a$ ($AU$)                     & $p$                               & $i$ ($^{\circ}$)                                      \\ \hline
\noalign{\smallskip}
   \citet{Heller2019} &  2020.3300$^{+0.0037}_{-0.0033}$ & 4.16959$^{+0.00051}_{-0.00053}$ & - & 0.0160$^{+0.0017}_{-0.0016}$ & - \\
   \noalign{\smallskip}  
   \citet{Kruse2019} & 1978.6338$^{+0.0062}_{-0.0056}$ & 4.16888$^{+0.00050}_{-0.00056}$ &  0.0503$^{+0.0227}_{-0.0114}$ & 0.0205$^{+0.0020}_{-0.0047}$ & - \\
   \noalign{\smallskip}  
   \texttt{EVEREST 2.0} & 1978.6357$^{+0.0024}_{-0.0023}$ & 4.1691$^{+0.0003}_{-0.0003}$ &     0.0371$^{+0.0026}_{-0.0020}$ & 0.0156$^{+0.0007}_{-0.0007}$ & 88.8116$^{+0.6143}_{-0.7022}$ \\
   \noalign{\smallskip}  
   \texttt{TFAW} & 1978.6359$^{+0.0011}_{-0.0012}$ & 4.1698$^{+0.0001}_{-0.0001}$ & 0.0350$^{+0.0009}_{-0.0010}$ & 0.0152$^{+0.0004}_{-0.0003}$ & 89.0010$^{+0.5117}_{-0.5574}$ \\
   \noalign{\smallskip}  
   \hline
   \noalign{\smallskip}  
   \multicolumn{6}{c}{95\% confidence intervals of the highest posterior density}\\ 
   \noalign{\smallskip}
   \hline    
   \noalign{\smallskip}
\noalign{\smallskip}  
\texttt{EVEREST 2.0} & 1978.62910 - 1978.64271 & 4.16828 - 4.16992 & 11.11698 - 15.22817 & 0.01355 - 0.01796 & 87.20658 - 89.99999 \\
\noalign{\smallskip}  
\texttt{TFAW} & 1978.63245 - 1978.63912 & 4.16946 - 4.17010 & 11.11686 - 12.73192 & 0.01424 - 0.01641 & 87.84871 - 90.00000 \\
\noalign{\smallskip}  
\hline
\noalign{\smallskip}  
\multicolumn{2}{c}{Derived system parameters} & \multicolumn{2}{c}{$R_{p}$ ($R_{\oplus}$)} & \multicolumn{2}{c}{b} \\ 
\noalign{\smallskip}  
\hline
   \noalign{\smallskip}  
\multicolumn{2}{c}{\citet{Heller2019}} & \multicolumn{2}{c}{1.10$^{+0.14}_{-0.12}$} & \multicolumn{2}{c}{0.36$^{+0.25}_{-0.24}$} \\
   \noalign{\smallskip}  
\multicolumn{2}{c}{\citet{Kruse2019}} & \multicolumn{2}{c}{1.41$^{+0.15}_{-0.34}$} & \multicolumn{2}{c}{-} \\
   \noalign{\smallskip}  
\multicolumn{2}{c}{\texttt{EVEREST 2.0}} & \multicolumn{2}{c}{1.072$^{+0.061}_{-0.054}$} & \multicolumn{2}{c}{0.26$^{+0.14}_{-0.16}$} \\
\noalign{\smallskip}
\multicolumn{2}{c}{\texttt{TFAW}} & \multicolumn{2}{c}{1.040$^{+0.037}_{-0.032}$} & \multicolumn{2}{c}{0.21$^{+0.11}_{-0.12}$} \\
\noalign{\smallskip}
   \hline                                   
   \end{tabular}
   \end{table*}

\begin{figure*}
\centering
\includegraphics[height=11.0cm,keepaspectratio]{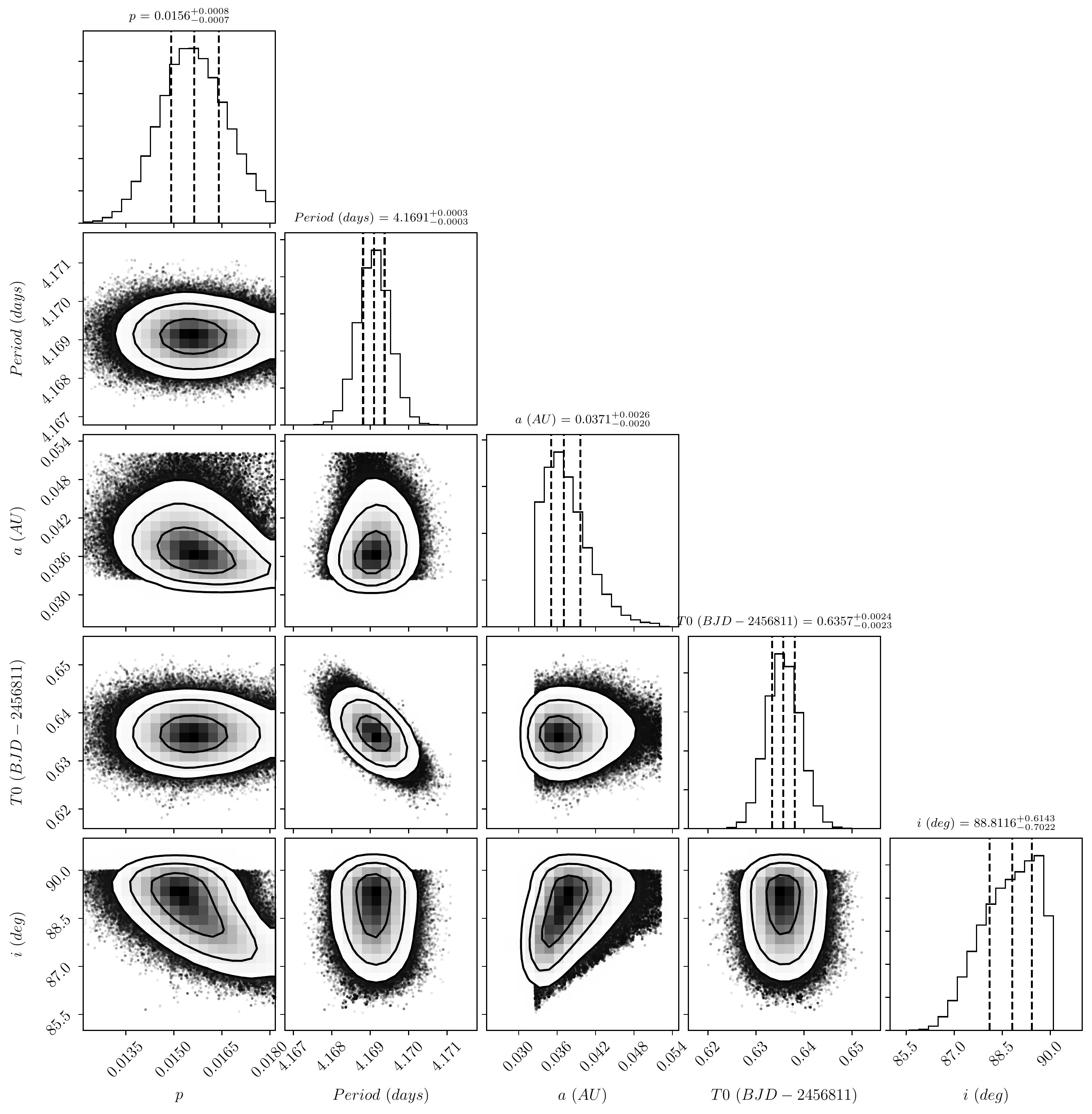}
\includegraphics[height=11.0cm,keepaspectratio]{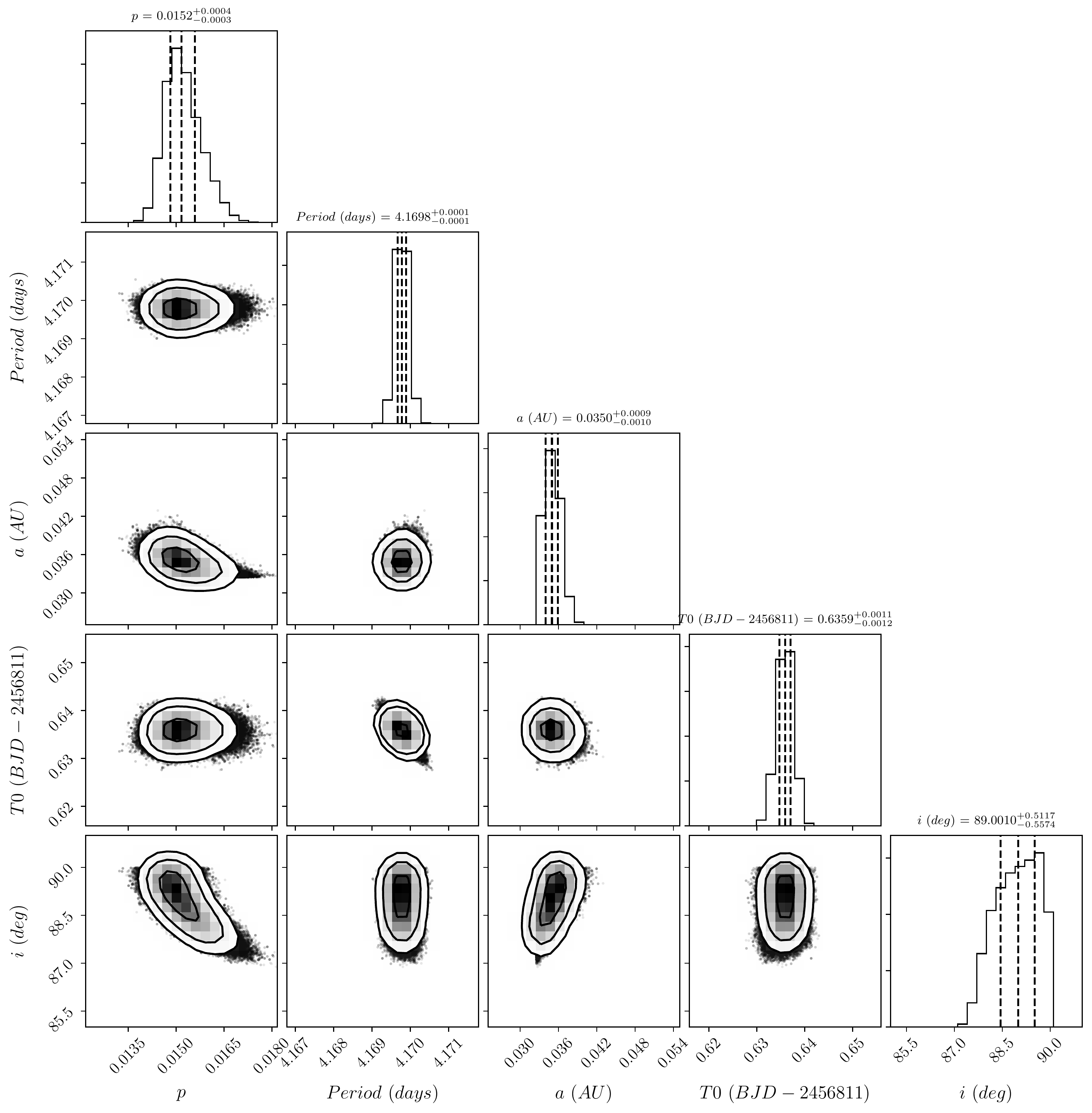}
\caption{1-D and 2-D projections of the posterior probability distributions of the 5 MCMC fitted parameters ($p$, $P$, $a$, $T_{0}$, $i$) for K2-298 b \texttt{EVEREST 2.0} (top) and \texttt{TFAW} (bottom) detrended light curves. The  25\%, 50\%, 75\% quantiles, are displayed in dashed vertical lines on the 1-D histograms and on the top each panels column.}
\label{fig:K2-298_b_mcmc_corner}
\end{figure*}

For the best fit (shown in Figure~\ref{fig:K2-298_b_mcmc_corner}), the mid-transit, $T_{0}$, for \texttt{EVEREST 2.0} and \texttt{TFAW} is compatible with the one reported by \citet{Kruse2019}, though for \texttt{TFAW}, the uncertainties are much lower than for the other two ($\sim$2$\times$ for \texttt{EVEREST 2.0}, and $\sim$5$\times$ compared to \citet{Kruse2019}). The value given by \citet{Heller2019} corresponds to the first transit after the center of the respective K2 target light curve. For the orbit inclination, $i$, neither \citet{Heller2019} nor \citet{Kruse2019} report a value. However, both \texttt{EVEREST 2.0} and \texttt{TFAW} yield values which are compatible within their errors and as with the previous parameters, \texttt{TFAW} returns the smallest uncertainties ($\sim$1.2$\times$ compared to \texttt{EVEREST 2.0}). For the period, $P$, the values found for \texttt{EVEREST 2.0} and \texttt{TFAW} are compatible within the errors with the one in \citet{Heller2019} and are very close to the one in \citet{Kruse2019}. Again, \texttt{TFAW} returns the smallest uncertainties for this parameter ($\sim$3$\times$ for \texttt{EVEREST 2.0}, $\sim$5$\times$ for \citet{Heller2019} and \citet{Kruse2019}). For the semi-major axis of the orbit, $a$, \texttt{EVEREST 2.0} and \texttt{TFAW} values are compatible taking the lower errors with the one reported by \citet{Kruse2019}. Again, \texttt{TFAW} returns the smallest uncertainties for this parameter ($\sim$2$\times$ for \texttt{EVEREST 2.0}, and $\sim$10$\times$ compared to \citet{Kruse2019}). Finally, regarding the planetary to star radius ratio, $p$, both \texttt{EVEREST 2.0} and \texttt{TFAW} obtain compatible values within the errors with the one reported by \citet{Heller2019}. With respect the value reported by \citet{Kruse2019}, it is compatible taking into account the reported planetary and stellar radius uncertainties. As with the other parameters, \texttt{TFAW} returns the smallest uncertainties for this parameter ($\sim$2$\times$ for \texttt{EVEREST 2.0}, $\sim$7$\times$ for \citet{Kruse2019}, and $\sim$5$\times$ compared to \citet{Heller2019}) and again, compatible with the $p/p_{0}$ dispersion seen in Figure \ref{fig:depth_bias}. As with K2-44 b, the widths of \texttt{TFAW}'s 95\% confidence intervals are narrower for all parameters.

With the best fit parameters, \texttt{EVEREST 2.0} obtains a mean planetary radius of $1.072^{+0.061}_{-0.054}$ $R_{\oplus}$ and \texttt{TFAW} $1.040^{+0.037}_{-0.032}$ $R_{\oplus}$, compatible with the $1.10^{+0.14}_{-0.12}$ $R_{\oplus}$ reported by \citet{Heller2019} and $1.41^{+0.15}_{-0.34}$ $R_{\oplus}$ reported by \citet{Kruse2019}. We also derive $b$ from the best fit parameters, for \texttt{EVEREST 2.0} to be $0.26^{+0.14}_{-0.16}$, and for \texttt{TFAW} to be $0.21^{+0.11}_{-0.12}$, both compatible with the one in \citet{Heller2019}. For both derived parameters, \texttt{TFAW} returns the lowest uncertainties.

\begin{figure*}
\centering
\includegraphics[height=15.0cm,keepaspectratio]{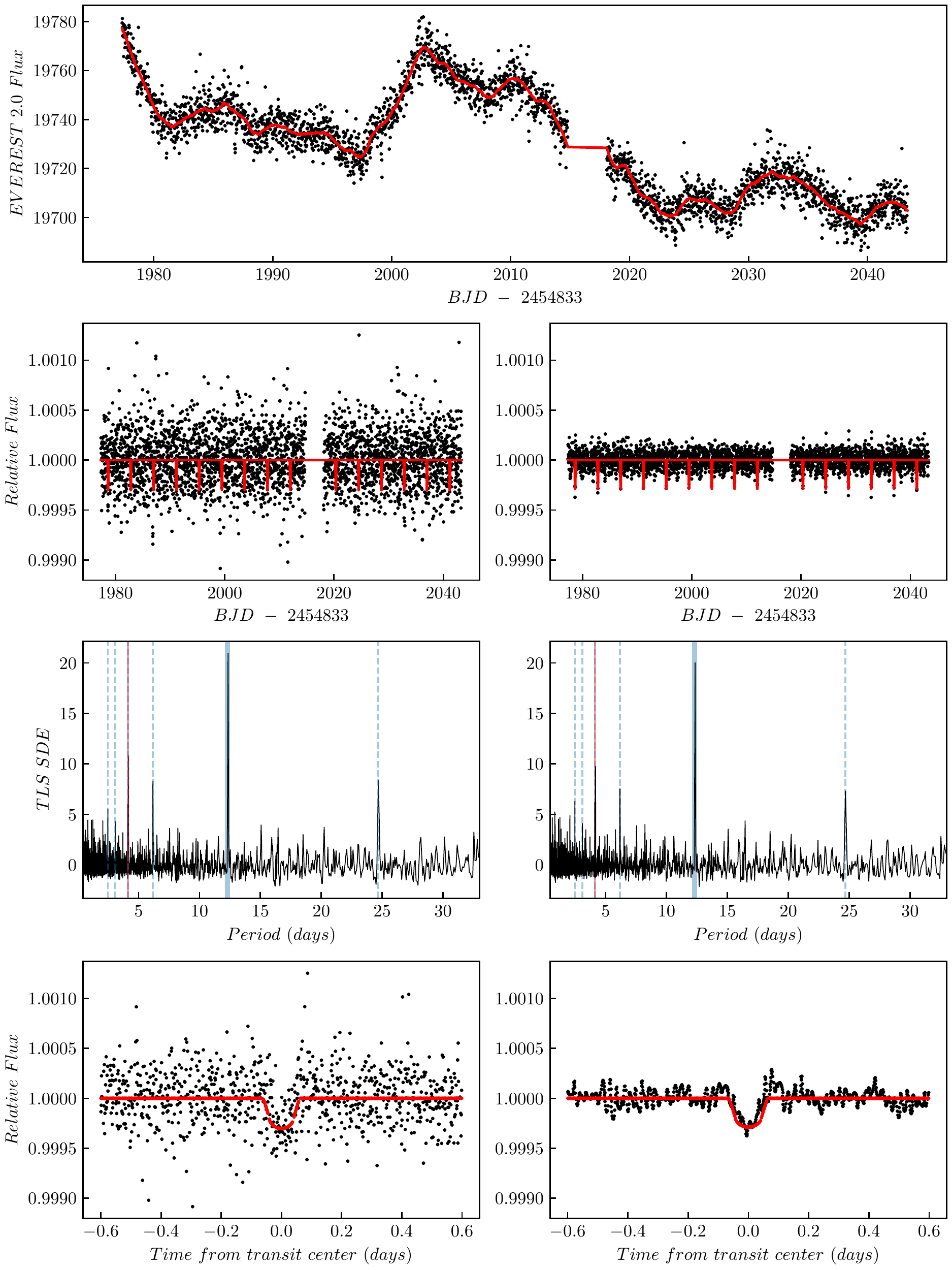}
\caption{\textit{From top to bottom}, the panels show the K2-298 raw flux of the \texttt{EVEREST 2.0} light curve (with the running median in red), the entire light curve with in-transit data of the new candidate marked in red (left for \texttt{EVEREST 2.0} and right for \texttt{TFAW}), the SDE$_{\rm TLS}$ periodogram (left for \texttt{EVEREST 2.0} and right for \texttt{TFAW}), 
and the normalized phase-folded to the K2-298 b period light curve with MCMC fit data marked in red (left for \texttt{EVEREST 2.0} and right for \texttt{TFAW}).}
\label{fig:k2-298b}
\end{figure*}

Figure~\ref{fig:k2-298b} shows the summary plot displaying the PLD, CBV-corrected flux provided by the \texttt{EVEREST 2.0} pipeline for K2-298 with the 2-day running median plotted in red, the \texttt{EVEREST 2.0} median-filtered light curve and the \texttt{TFAW}-corrected light curves (with the MCMC derived transit data of the new candidate plotted in red), the \texttt{TLS} periodograms for \texttt{EVEREST 2.0} and \texttt{TFAW}'s frequency analysis step; and the phase-folded light curves with the MCMC fit data (red line) for \texttt{EVEREST 2.0} (left) and \texttt{TFAW} iteratively denoised and reconstructed one (right). The \texttt{TLS} periodograms show the position of the candidate planet detected period \citep{Kruse2019} (solid blue line) and its harmonics (dashed blue lines) and the position of K2-298 b period (solid red line).

\subsubsection{\texttt{TFAW}'s limitations}
\label{subsec:limitations}
The small oscillations in some of the \texttt{TFAW} detrended and denoised light curves, such as in the lower right panel of Figure~\ref{fig:k2-44b}, might be explained by one of or a combination of the following three factors. First, real correlated noise in the decomposition levels above the chosen signal level that may have not been properly removed during the \texttt{TFAW} detrending stage. Second, in some cases, real stellar variability with different time scales can be present at the same decomposition levels associated with the phased folded transit signal. As a consequence, this stellar variability can be included in the signal estimation, reconstruction and denoising process. Third, spurious features in the signal estimation arising from the shape of the mother wavelet or an alias due to the sampling of the data in the phase folded light curve. These two later effects cannot be fully discarded but the signal estimation can be improved by increasing the number of decomposition levels or by optimizing the signal level selection criteria. On the other hand, signals caused by stellar activity or pulsation are unique to each star and temporally correlated, and can not be easily removed by decorrelation or denoising techniques. For example, the \textit{bump} around phase [+0.2,+0.3] seen in the K2-44 b \texttt{TFAW} phase-folded light curve at the lower right panel of Figure~\ref{fig:k2-44b}, is also present in the \texttt{EVEREST 2.0} at the same phase interval.
The removal of temporally correlated noise can be difficult to do using the SWT. Although this is beyond the scope of this paper, future versions of the \texttt{TFAW} can benefit from: the incorporation of GP to model the covariance structure of the correlated noise \citep{Chakrabarty2019}, modeling of the signal of interest in the wavelet domain \citep{Goossens2009} or thresholding the wavelet coefficients \citep{Jansen1999} prior to the Inverse Stationary Wavelet Transform (ISWT) at each level to reduce the correlated noise. Another way to diminish the effects of correlated noise (which has a higher frequency in the phase folded light curve than the transit signal) would be to increase the noise level or to increase the number of decomposition levels (i.e. adding more epochs to the light curve) as to better separate the different signal contributions at different frequencies. The later is not possible with K2 data as the number of available epochs in the archive is fixed. It is worth mentioning that we have been extremely careful when selecting the noise and signal levels to minimize the chances of removing part of the signal of interest contribution, so the results could improve by a more aggressive selection of those levels.

\section{Testing transit search with \texttt{TFAW}: two transit candidates in K2 observing campaign C1}
\label{sec:k2c1}

In this Section we present partial results obtained after the application of \texttt{TFAW} to \texttt{EVEREST 2.0} light curves from the K2 observing campaign C1. We present two new Earth-sized transiting planet candidates detected using \texttt{TLS} during \texttt{TFAW}'s frequency analysis step (i.e. light curve is detrended and has had a first SWT noise estimation removed). For both cases we show their transit parameters obtained with MCMC after \texttt{TFAW}'s iterative signal denosing and reconstruction. A more extensive planet search, not only for C1 but for all K2 campaigns, and using fully automatic vetting \citep{Kostov2019,Zink2020} is underway. With this study, which is to be completed soon for an upcoming publication, we will obtain more potential new candidates. 

In this Section we also compare our ability to recover confirmed and candidate transit planets detected by other searches for C1.

\subsection{Data description}

We use the PLD, CBV-corrected fluxes provided by the \texttt{EVEREST 2.0} pipeline for K2 observing campaign C1. For each light curve, 3072 epochs with the \texttt{QUALITY}=0 flag are considered. Given that the goal is to search for transiting signals, a 2-day Savitsky-Golay filter is applied to remove or minimize the effects of stellar variability. After the filter has been applied, light curves have their outliers removed using a 5$\sigma$ clipping. An extra outlier removal is done by \texttt{TFAW} prior to the period search using a wavelet estimation of the signal (see \citet{delSer2018} for more details). To iteratively denoise and reconstruct the light curves with \texttt{TFAW}, a template of reference stars for each light curve ($\sim$90) is built from a sub-set of stars from the same CCD module. To avoid the inclusion of variable stars in this template, we use Stetson's L variability index \citep{Stetson1996}.

\subsection{Transit search and vetting criteria}
\label{subsect:vetting}

We follow a transit search, vetting, and False-Positive Probability (FPP) approach similar to the one detailed in \citet{Heller2019}. First, we use \texttt{TLS} to search for transiting signals during \texttt{TFAW}'s frequency analysis step. \texttt{TLS} is run using modelled stellar parameters, $M_{\rm s}$\,, $R_{\rm s}$\,, $u_{\rm 1}$\,, and $u_{\rm 2}$\,. All light curves that have a \texttt{TLS} power spectrum peak with an SDE$_{\rm TLS}$ greater than 9 (false-positive rate $<$10$^{-4}$ \citep{Hippke2019}) are considered a significant detection and undergo \texttt{TFAW}'s iterative signal reconstruction and denoising. Then, we visually inspect those \texttt{TFAW}-corrected light curves and only keep the ones which visually show transit-like features.

Following the same procedure as \citet{Heller2019}, all transits are required to have at least 0.5 days from the beginning or end of any gaps in their light curves to avoid false positives. If a candidate had three or fewer transits, they should have a SNR$>$10. In addition, to reject eclipsing binaries, for all candidates the average depth of the odd and even transits should agree within $<$3$\sigma$ and objects should not present evidence of a secondary eclipse at the $>$3$\sigma$ level at half an orbital phase after the candidate transit. 

For those light curves we consider of interest, we include some extra vetting steps with respect to the procedure in \citet{Heller2019} to increase the reliability of the candidates. First, we perform an extensive bibliographic search of such target. We cross-match this source with the most up-to-date (8 Mar 2020 for this work) K2-C1 lists of confirmed or candidate exoplanets from the NASA Exoplanet Archive\footnote{\url{https://exoplanetarchive.ipac.caltech.edu}} \citep{Akeson2013} or in the Vizier database. We also check if the target shows any known kind of variability or pulsation \citep{Armstrong2015,Armstrong2016,Watson2006}. The updated stellar parameters and the 2MASS and SDSS photometry of the host stars are retrieved from catalogs such as EPIC \citep{Huber2016}, Gaia-DR2 \citep{Gaia2018} and \citet{Claret2018}. We also search if the candidates have been observed by the \emph{TESS} mission \citep{Ricker2015}. Second, we rule out that no other light curve in the same CCD module present transit-like features with similar periods and transit epochs as the candidate. Third, we compare the \texttt{TLS} and \texttt{BLS} \citep{Kovacs2002} periodograms to check if they present similar peaks. Next, to diminish the chances of the \texttt{TLS} detection being fortuitous, we randomly reshuffle the target light curve to remove the transiting signal and to simulate 1,000 light curves with similar CDPP as the \texttt{TFAW}'s frequency analysis step light curve. Using the \texttt{batman} package, we inject a transit signal with the parameters found by \texttt{TLS}. We run \texttt{TLS} over the simulated light curves and check whether the transit is recovered or not. If the transit is recovered in more than 90\% of the simulated light curves, then it passes to the final vetting step. We use the publicly available \texttt{vespa}\footnote{\url{https://github.com/timothydmorton/VESPA}} software \citep{Morton2012,Morton2015vespa} to evaluate the FPP of our transit candidates. For each of the candidates, we supply the software with their corresponding \texttt{TFAW} phase folded light curve, their celestial coordinates, the stellar parameters of their host star, and its photometry. We also compute a limiting aperture obtained from the validation sheets from the \texttt{EVEREST 2.0} database, and inspect independent photometry and high angular resolution images to evaluate contamination from other sources. Using such information, \texttt{vespa} calculates the probabilities of the transiting signal being caused by non-associated blended eclipsing binaries, eclipsing binaries, hierarchical triples and non-associated stars with transiting planets. Only candidates with a FPP lower than 1\% are considered as valid candidates.

\subsection{Comparison with other searches}
\label{subsect:recoverknowntransits}

In Table \ref{tab:catdetect} we show that our K2-C1 \texttt{TFAW} and \texttt{TLS} based planet search is able to recover all confirmed planets and most of the K2 candidates from previous studies compiled in NASA Exoplanet Archive. Generally, the missing ones are usually single-transit events, not suitable for periodic signal searching algorithms like \texttt{TLS}, multi-periodic systems for which the current \texttt{TFAW} version only reconstructs the signal for the most significant period (though the other planets in the system might have also been detected in the \texttt{TLS} periodogram but with lower SDEs), or some that present an SDE$_{\rm TLS}<$9 (usually above 6.5) but have their most significant peak at the catalogued period. For the confirmed planets, \texttt{TFAW} finds at least one planet for each of the 35 cataloged planetary systems. For the ones in \citet{Barros2016}, \texttt{TFAW} detects 18 transiting systems, one, with a period $>$40 days is missed by \texttt{TLS}, and for another \texttt{TLS} does not find a significant peak at the listed period. For \citet{Crossfield2016}, we detect 9 of the 13 listed candidates for C1. Two of the missing ones are candidates with periods $>$40 days; for the other two \texttt{TLS} does not find a significant peak at the listed periods. Regarding \citet{Vanderburg2016}, we detect 68 of the listed candidates. The other four are single-transit or have a period $>$40 days. We detect all the candidates for C1 listed by \citet{Mayo2018}. Finally, we detect at least one planet in all 78 non-single-transit systems in \citet{Kruse2019}. We believe our new candidates presented here went undetected by other groups due to the combination of two factors, the increased photometric precision achieved with \texttt{TFAW}, specially for faint magnitudes, together with \texttt{TLS} improved capabilities to detect smaller planets.

\begin{table}
\centering
\caption{Comparison of our planet search to previous groups, all sub-sampled to K2-C1 campaign.}
\label{tab:catdetect}
\begin{tabular}{l c c}
\hline
Candidate List  & Cataloged planets & Number we found  \\
\hline
Confirmed & 48 & 35 \\
\citet{Barros2016} & 20 & 18 \\
\citet{Crossfield2016} & 13 & 9 \\ 
\citet{Vanderburg2016} & 72 & 68 \\
\citet{Mayo2018} & 13  & 13 \\ 
\citet{Kruse2019} & 97 & 78 \\ 
\hline
\end{tabular}
\end{table}

\subsection{Two new transit candidates from K2-C1 data}
\label{subsect:detectionnewtransits}

In this Section we present two new transit candidates detected using the combination of the increased photometric precision of \texttt{TFAW}-corrected light curves and \texttt{TLS}. Both candidates have passed the vetting procedure explained in \ref{subsect:vetting}. 
As with K2-44 b and K2-298 b, once detected, we checked the \texttt{EVEREST 2.0} light curves after masking the new candidate transits (given their smaller transit depths, the effect of the PLD can decrease the transit depth up to $\sim$10\%~\citep{Luger2018}). Finally, in order to determine the transit parameters of these new candidates, we use the \texttt{TFAW}-corrected light curves, the \texttt{TLS} output and the cataloged stellar properties as the starting point for the MCMC fit. As mentioned before, a complete study of observing campaign C1 (and C2-C18) is underway where more transit candidates are expected to be found. In Tables \ref{tab:EPIC201170410.02} and \ref{tab:EPIC201757695.02} we summarize the stellar and transit properties of the two new planetary candidates fully described in Sections~\ref{subsubsect:EPIC201170410} and~\ref{subsubsect:EPIC201757695}.

\subsubsection{EPIC 201170410}
\label{subsubsect:EPIC201170410}

EPIC 201170410 is a $K_p$=15.673 mag, $G$=16.4386 mag \citep{Gaia2018}, $K_s$=12.619$\pm$0.027 mag, $(J-K_{s})$=0.84$\pm$0.037 \citep{Cutri2003} star. It is located at $(\alpha,~\delta)$ = (11:20:33.81, -04:48:25.21) \citep{Gaia2018} and observed by the K2 mission during the C1 monitoring campaign, channel 55.

Table \ref{tab:EPIC201170410.02} lists the cataloged stellar parameters by \citet{Huber2016} and \citet{Stassun2019} for this source. \citet{Gaia2018} provides a $T_{eff}$=$4013^{+714}_{-713}$ K, which is compatible within the previous listed value. This target is classified as a dwarf by  \citet{Stassun2019} and, given its cataloged radius, mass and effective temperature it is most likely an M-dwarf. No variability detection for this star is provided either in \citet{Armstrong2015,Armstrong2016}, \citet{Watson2006} or \citet{Gaia2018}.

\begin{table}
\centering
\caption{Stellar parameters for EPIC 201170410 host star, \texttt{TLS}-detection parameters for the transit planetary candidate EPIC 201170410.02, \texttt{TLS} and \texttt{vespa} vetting parameters, and MCMC fit and derived planetary candidate parameters of EPIC 201170410.02.} 
\label{tab:EPIC201170410.02}
\begin{tabular}{ll}
\hline
K2 ID & EPIC 201170410 \\ 
\hline
Stellar parameters &  \\
Stellar radius $R_{\rm s}$ ($R_{\odot}$) & $0.282^{+0.074}_{-0.069}$ $^{\star}$ \\
Stellar mass $M_{\rm s}$ ($M_{\odot}$) & $0.287^{+0.101}_{-0.084}$ $^{\star}$ \\
Effective temperature (K) & $3648^{+172}_{-143}$ $^{\star}$ \\
Surface gravity ($log_{10}(cm/s^2$)) & 4.999$\pm$0.075 $^{\star}$ \\
Metallicity [Fe/H] & $-0.048^{+0.150}_{-0.210}$ $^{\star}$ \\
Distance (pc) & $134.0^{+45.8}_{-39.8}$ $^{\star}$ \\
Luminosity $L_{\rm s}$ ($L_{\odot}$) & - \\
Luminosity class & Dwarf $^{\star\star}$ \\
Spectral Type & - \\ 
\hline
\texttt{TLS}-detection transit parameters &  \\
Number of transits & 10 $^{\dagger}$ \\
Period $P$ (days) & 6.799025 \\
Epoch $T_0$ (BJD - 2454833) (days) & 1980.14697625 \\
Duration (hours) & 1.84933104 \\
Depth $\delta$ (\%) & 0.136 \\
SNR & 6.97 $^{\dagger\dagger}$ \\
Radius ratio $p$ & 0.0335913 \\
Scaled semi-major axis $a$ (AU) & 0.0463297 \\
Planetary radius $R_{p}$ ($R_{\oplus}$) & 1.0332548 \\
\hline
\texttt{TLS} and \texttt{vespa} vetting parameters &  \\
SDE$_{\rm TLS}$ & 9.382 \\
$\rho\,($\arcsec$)$ & 22.5 \\
FPP & $4.8~\times~10^{-10}$ \\
\hline
MCMC transit parameters &  \\
Period $P$ (days) & 6.7987$\pm$0.0001 \\
Epoch $T_0$ (BJD - 2454833) (days) & 1980.1485$^{+0.0006}_{-0.0005}$ \\
Eccentricity $e$ & 0 (assumed) \\
Radius ratio $p$ & 0.0340$^{+0.0007}_{-0.0006}$ \\
Scaled semi-major axis $a$ (AU) & $0.0349^{+0.0027}_{-0.0022}$ \\
Inclination $i$ ($^{\circ}$) & $89.0025^{+0.4142}_{-0.2885}$ \\
\hline
Derived planetary parameters &  \\
Planetary radius $R_{p}$ ($R_{\oplus}$) & $1.047^{+0.276}_{-0.257}$ $^{\ddagger}$ \\
Impact parameter $b$ & $0.46^{+0.20}_{-0.14}$ $^{\ddagger}$ \\
\hline
\multicolumn{2}{l}{$^{\dagger}$ denotes number of transits detected by \texttt{TLS} that have data.} \\ 
\multicolumn{2}{l}{$^{\ddagger}$ denotes values derived from fitted values.} \\ 
\multicolumn{2}{l}{$^{\dagger\dagger}$ as defined in \citet{Pont2006}.} \\ 
\multicolumn{2}{l}{$^{\star}$ denotes values derived from \citet{Huber2016}.}\\
\multicolumn{2}{l}{$^{\star\star}$ denotes values derived from \citet{Stassun2019}.}\\
\end{tabular}
\end{table}

Table \ref{tab:EPIC201170410.02} also summarizes the detection and vetting parameters obtained with \texttt{TLS} and \texttt{vespa} for the EPIC 201170410 \texttt{TFAW}-corrected light curve. With an SDE$_{\rm TLS}$>9.0 and an FPP of $4.8~\times~10^{-10}$ it can be considered an statistically validated exoplanet candidate \citep{Heller2019}. Finally, it also lists the transit and derived planetary parameters obtained from the MCMC fit of the \texttt{TFAW}-corrected light curve.

\begin{figure*}
\centering
\includegraphics[height=15.0cm,keepaspectratio]{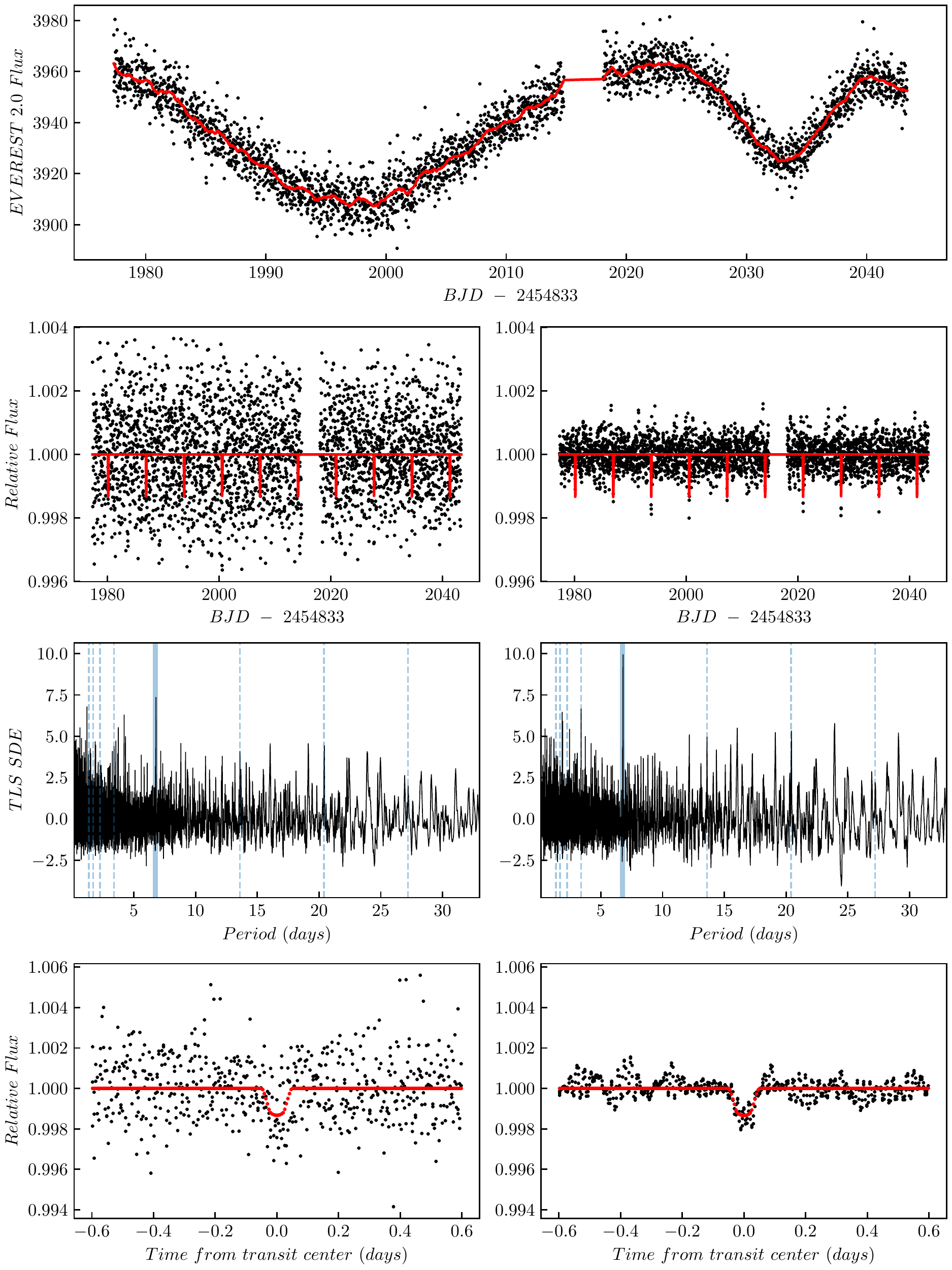}
\caption{\textit{From top to bottom}, the panels show the EPIC 201170410 raw flux of the \texttt{EVEREST 2.0} light curve (with the running median in red), the entire light curve with in-transit data of the new candidate marked in red (left for \texttt{EVEREST 2.0} and right for \texttt{TFAW}), the SDE$_{\rm TLS}$ periodogram (left for \texttt{EVEREST 2.0} and right for \texttt{TFAW}), and the normalized phase-folded light curve with MCMC fit data marked in red (left for \texttt{EVEREST 2.0} and right for \texttt{TFAW}).}
\label{fig:epic201170410}
\end{figure*}

Figure~\ref{fig:epic201170410} shows the summary plot displaying the PLD, CBV-corrected flux provided by the \texttt{EVEREST 2.0} pipeline for EPIC 201170410 with the 2-day running median plotted in red, the \texttt{EVEREST 2.0} median-filtered light curve and the \texttt{TFAW}-corrected light curves (with the MCMC derived transit data of the new candidate plotted in red), the \texttt{TLS} periodograms for \texttt{EVEREST 2.0} and \texttt{TFAW}'s frequency analysis step; and the phase-folded lights curve with the MCMC fit data (red line) for \texttt{EVEREST 2.0} (left) and \texttt{TFAW} iteratively denoised and reconstructed one (right). The \texttt{TLS} periodograms show the position of the candidate's detected period (solid blue line) and its harmonics (dashed blue lines).

As can be seen from the \texttt{TLS} periodograms, the peak at 6.7987 days is visible both for \texttt{EVEREST 2.0} and \texttt{TFAW}. However, for the latter, and given \texttt{TFAW}'s denoising capabilities, this peak has a higher SDE$_{\rm TLS}$ (9.4 versus 7.4) that allows it to cross our significant detection threshold (i.e $>$9). In the particular case of \texttt{EVEREST 2.0} detrended light curve, EPIC 201170410.02 might not have been detected by other authors, even those using \texttt{TLS}, because the light curve noise sets the SDE$_{\rm TLS}$ below the detection threshold employed by them (9.0 in the case of \citet{Hippke2019} and \citet{Heller2019}). In addition, the transit-like feature becomes clearly visible in the reconstructed \texttt{TFAW} light curve (right plot in the second row in Figure~\ref{fig:epic201170410} and right bottom plot in Figure~\ref{fig:epic201170410}). It is worth mentioning that the wavelet does not have an a priori knowledge of the shape of the underlying signal in a given light curve and that its shape is estimated only from the phase folded light curve at a given period and is re-estimated at every iteration step during \texttt{TFAW}'s signal reconstruction step. As show in Section~\ref{subsec:k2-298} with K2-298 b, EPIC 201170410 shows another example of the potential of \texttt{TFAW} to detect and recover transit-like features even for highly noisy light curves as this one.

To confirm that our candidate light curve was not affected by contamination of nearby sources, we visually inspect the K2-C1, channel 55, calibrated full frame images (FFI) for EPIC 201170410. In addition, \citet{Gaia2018} does not list any other source within EPIC 201170410 aperture either. As explained in Section~\ref{subsect:vetting}, using \texttt{EVEREST 2.0} validation sheets for EPIC 201170410 we assigned the maximum aperture radius for \texttt{vespa}, $\rho\,$=22.5$\arcsec$, to estimate the possibility of contamination by other objects.

EPIC 201170410 has also been observed by the \emph{TESS} mission in Sector 9 (28 Feb to 26 Mar 2019). EPIC 201170410 corresponds to TIC 38116202. However, no light curve for this target is available from the usual \emph{TESS} data archives. We download the \emph{TESS} FFI corresponding to TIC 38116202 using the \texttt{eleanor}\footnote{\url{https://github.com/afeinstein20/eleanor}} package and use it to compute raw, CBV-corrected and PSF-modeled light curves for this target. TIC 38116202 is highly affected by crowding from two other near and brighter stars in the \texttt{TESS} FFI leading to a worse photometric solution than the one from K2 thus, no conclusive results can be obtained from this light curve after running the period search using \texttt{TLS}.

Finally, in order to better determine the EPIC 201170410.02 transit and planetary parameters, we run a MCMC fit over the \texttt{TFAW}-corrected light curve using the \texttt{TLS} output as initial guess values. Table \ref{tab:EPIC201170410.02} lists the obtained MCMC fit and planetary derived values. For EPIC 201170410.02 and, assuming a circular orbit, \texttt{TFAW} yields a 1.047$^{+0.276}_{-0.257}R_{\oplus}$ planet orbiting its host star at 0.0349$^{+0.0027}_{-0.0022}$ AU with a period of 6.7987$\pm0.0001$ days. With this radius value, EPIC 201170410.02 is the 10-th smallest K2-C1 planet within those candidates with an estimated planetary radius and the 44-th smallest in all K2 campaigns.

\subsubsection{EPIC 201757695}
\label{subsubsect:EPIC201757695}

EPIC 201757695 is a $K_p$=14.599 mag, $G$=14.6206 mag \citep{Gaia2018}, $K_s$=12.149$\pm$0.026 mag, $(J-K_{s})$=0.707$\pm$0.037 \citep{Cutri2003} star. It is located at $(\alpha,~\delta)$ = (11:35:45.24, +04:36:59.21) \citep{Gaia2018} and observed by the K2 mission during the C1 monitoring campaign, channel 47. 

Table \ref{tab:EPIC201757695.02} lists the cataloged stellar parameters by \citet{Huber2016}, \citet{Bailer2018}, \citet{Gaia2018} and \citet{Stassun2019} for this source. In addition, \citet{Gaia2018} provides a $T_{eff}$=$4706^{+290}_{-264}$ K, which is compatible within the previous listed value. This star is classified as a dwarf \citep{Stassun2019} and, given its cataloged radius, mass and effective temperature it is most likely a K-type star. No variability detection for this star is provided either in \citet{Armstrong2015,Armstrong2016} or \citet{Gaia2018}. It is cataloged as a "variable star of unspecified type" (VAR) by the AAVSO International Variable Star Index (VSX)\footnote{\url{https://www.aavso.org/vsx/index.php}} \citep{Watson2006} with a period of 19.94041 days.

\begin{table}
\centering
\caption{Stellar parameters for EPIC201757695 host star, \texttt{TLS}-detection parameters for the  transit planetary candidate EPIC201757695.02, \texttt{TLS} and \texttt{vespa} vetting parameters, and MCMC fit and derived planetary candidate parameters of EPIC201757695.02.} 
\label{tab:EPIC201757695.02}
\begin{tabular}{ll}
\hline
K2 ID & EPIC 201757695 \\ 
\hline
Stellar parameters &  \\
Stellar radius $R_{\rm s}$ ($R_{\odot}$) & $0.655^{+0.041}_{-0.045}$ $^{\star}$ \\
Stellar mass $M_{\rm s}$ ($M_{\odot}$) & $0.727^{+0.044}_{-0.053}$ $^{\star}$ \\
Effective temperature (K) & $4520^{+108}_{-54}$ $^{\star}$ \\
Surface gravity ($log_{10}(cm/s^2$)) & $4.659^{+0.035}_{-0.025}$ $^{\star}$ \\
Metallicity [Fe/H] & $-0.003^{+0.120}_{-0.300}$ $^{\star}$ \\
Distance (pc) & $577.6^{+33.2}_{-29.8}$ $^{\star\star}$ \\
Luminosity $L_{\rm s}$ ($L_{\odot}$) & $0.433^{+0.482}_{-0.385}$ $^{\star\star\star}$ \\
Luminosity class & Dwarf $^{\star\star\star\star}$ \\
Spectral Type & - \\ 
\hline
\texttt{TLS}-detection transit parameters &  \\
Number of transits & 30 $^{\dagger}$ \\
Period $P$ (days) & 2.04779036 \\
Epoch $T_0$ (BJD - 2454833) (days) & 1978.6419269 \\
Duration (hours) & 1.82256 \\
Depth $\delta$ (\%) & 0.021 \\
SNR & 11.04 $^{\dagger\dagger}$ \\
Radius ratio $p$ & 0.01248 \\
Scaled semi-major axis $a$ (AU) & 0.0283772 \\
Planetary radius $R_{p}$ ($R_{\oplus}$) & 0.891695 \\
\hline
\texttt{TLS} and \texttt{vespa} vetting parameters &  \\
SDE$_{\rm TLS}$ & 15.005 \\
$\rho\,($\arcsec$)$ & 18.3 \\
FPP & $8.13~\times~10^{-5}$ \\
\hline
MCMC transit parameters &  \\
Period $P$ (days) & 2.0478$\pm$0.0001 \\
Epoch $T_0$ (BJD - 2454833) (days) & 1978.6370$^{+0.0008}_{-0.0007}$ \\
Eccentricity $e$ & 0 (assumed) \\
Radius ratio $p$ & 0.0127$\pm$0.0002 \\
Scaled semi-major axis $a$ (AU) & 0.0296$\pm$0.0005 \\
Inclination $i$ ($^{\circ}$) & $89.2757^{+0.3689}_{-0.3717}$ \\
\hline
Derived planetary parameters &  \\
Planetary radius $R_{p}$ ($R_{\oplus}$) & $0.90819^{+0.05861}_{-0.06401}$ $^{\ddagger}$ \\
Impact parameter $b$ & 0.12$\pm$0.06 $^{\ddagger}$ \\
\hline
\multicolumn{2}{l}{$^{\dagger}$ denotes number of transits detected by \texttt{TLS} that have data.} \\ 
\multicolumn{2}{l}{$^{\ddagger}$ denotes values derived from fitted values.} \\ 
\multicolumn{2}{l}{$^{\dagger\dagger}$ as defined in \citet{Pont2006}.} \\ 
\multicolumn{2}{l}{$^{\star}$ denotes values derived from \citet{Huber2016}.}\\
\multicolumn{2}{l}{$^{\star\star}$ denotes values derived from \citet{Bailer2018}.}\\
\multicolumn{2}{l}{$^{\star\star\star}$ denotes values derived from \citet{Gaia2018}.}\\
\multicolumn{2}{l}{$^{\star\star\star\star}$ denotes values derived from \citet{Stassun2019}.}\\
\end{tabular}
\end{table}

Table \ref{tab:EPIC201757695.02} also lists the detection and vetting parameters obtained with the \texttt{TLS} and \texttt{vespa} for the EPIC 201757695 \texttt{TFAW}-corrected light curve. With an SDE$_{\rm TLS}>$9 and an FPP of 8.13$\times10^{-5}$ EPIC 201757695.02 can be considered an statistically validated exoplanet candidate. Finally, Table \ref{tab:EPIC201757695.02} also summarizes the transit and derived planetary parameters obtained from the MCMC fit of the \texttt{TFAW}-corrected light curve.

\begin{figure*}
\centering
\includegraphics[height=15.0cm,keepaspectratio]{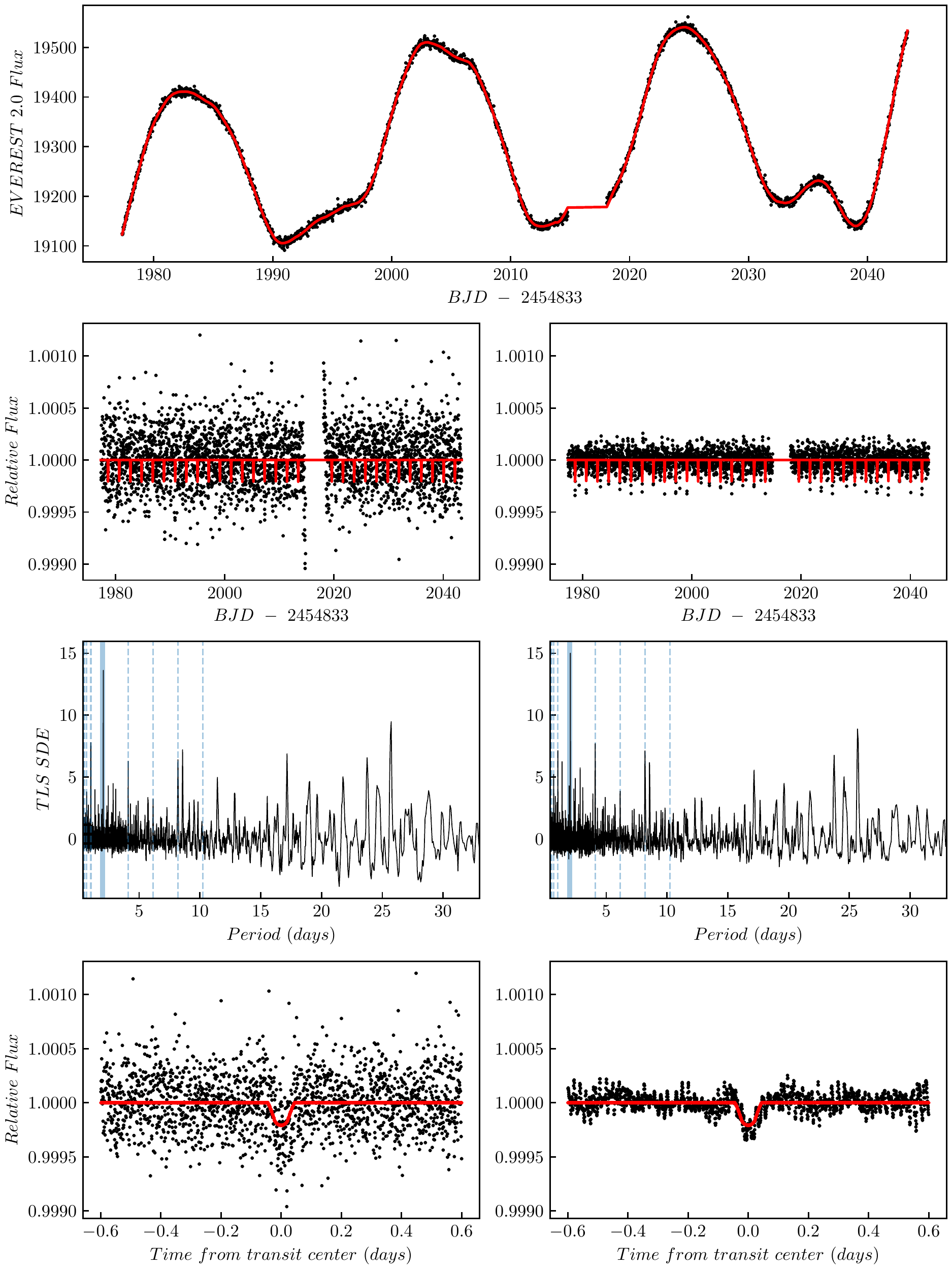}
\caption{\textit{From top to bottom}, the panels show the EPIC 201757695 raw flux of the \texttt{EVEREST 2.0} light curve (with the running median in red), the entire light curve with in-transit data of the new candidate marked in red (left for \texttt{EVEREST 2.0} and right for \texttt{TFAW}), the SDE$_{\rm TLS}$ periodogram (left for \texttt{EVEREST 2.0} and right for \texttt{TFAW}), and the normalized phase-folded light curve with MCMC fit data marked in red (left for \texttt{EVEREST 2.0} and right for \texttt{TFAW}).}
\label{fig:epic201757695}
\end{figure*}

Figure~\ref{fig:epic201757695} shows the summary plot displaying the PLD, CBV-corrected flux provided by the \texttt{EVEREST 2.0} pipeline for EPIC 201757695 with the 2-day running median plotted in red, the \texttt{EVEREST 2.0} median-filtered light curve and the \texttt{TFAW}-corrected light curves (with the MCMC derived transit data of the new candidate plotted in red), the \texttt{TLS} periodograms for \texttt{EVEREST 2.0} and \texttt{TFAW}'s frequency analysis step; and the phase-folded light curves with the MCMC fit data (red line) for \texttt{EVEREST 2.0} (left) and \texttt{TFAW} iteratively denoised and reconstructed one (right). The \texttt{TLS} periodograms show the position of the candidate's detected period (solid blue line) and its harmonics (dashed blue lines).

The \texttt{TLS} periodograms in Figure \ref{fig:epic201757695} show a clear peak at 2.048 days and with enough  SDE$_{\rm TLS}$ to cross our detection threshold both for \texttt{EVEREST 2.0} and \texttt{TFAW}. However, for the latter, again due to \texttt{TFAW} denoising capabilities, this significance is higher (15.0 versus 14.2). As with the case of EPIC 201170410, for EPIC 201757695 the transit feature becomes clearly more visible in the \texttt{TFAW}-corrected light curve.

To confirm that this second candidate was not affected by contamination from nearby sources, we visually inspect the K2 calibrated full frame images (FFI) for EPIC 201757695. In addition, \citet{Gaia2018} does not show any other source within EPIC 201757695 aperture. Using \texttt{EVEREST 2.0} validation sheets for this target, we assigned the maximum aperture radius for \texttt{vespa}, $\rho\,$=18.3$\arcsec$ as listed in Table~\ref{tab:EPIC201757695.02}.

EPIC 201757695 is included in the \emph{TESS} Input Catalog \citep{Stassun2019} as TIC 903075188. However, to date 31 Mar 2020, it has not been observed yet and no light curve is available for this target.

Again, in order to better determine EPIC 201757695.02 transit and planetary parameters, we run a MCMC fit over this target \texttt{TFAW}-corrected light curve using the \texttt{TLS} output as initial guess values.

Table~\ref{tab:EPIC201757695.02} lists the obtained MCMC fit and planetary derived values. For EPIC 201757695.02 and, assuming a circular orbit, \texttt{TFAW} yields a 0.908$^{+0.059}_{-0.064}R_{\oplus}$ planet orbiting its host star at a distance of 0.0296$\pm0.0005$ AU with a period of 2.0478$\pm0.0001$ days. With this radius value, EPIC 201757695.02 is the 9-th smallest K2-C1 of the candidates with an estimated planetary radius, and the 39-th smallest in all K2 campaigns.

\section{Conclusions}

We present the results obtained after applying the wavelet-based \texttt{TFAW} to further extend the photometric precision achieved by \texttt{EVEREST 2.0} light curves from the K2 mission. We compare the photometric precision of both algorithms in terms of the 6hr CDPP. On average, the \texttt{TFAW} median 6hr CDPP is $\sim$30$\%$ better than the one from \texttt{EVEREST 2.0}. This improvement can reach about $\sim$35-40$\%$ in the faint magnitude range during \texttt{TFAW}'s frequency analysis step. The 6hr CDPP of \texttt{TFAW} iteratively reconstructed and denoised light curves (i.e. those that have crossed the signal detection threshold) can be $\sim$50-75$\%$ better than the corresponding \texttt{EVEREST 2.0} one. We show that the transit detection efficiency of simulated Earth-Sun-like systems along the 8$< K_p <$18 magnitude range for \texttt{TFAW} is a factor $\sim$8.5$\times$ on average higher than for the case of \texttt{EVEREST 2.0} light curves. This improvement increases up to $\sim$21$\times$ for the faint ($K_p >15.0$) magnitude range. We validate our algorithm by performing transit injection/recovery tests where the planetary radius to stellar radius ratios (i.e transit depths) are recovered without significant bias in the \texttt{TFAW}-corrected light curves.

We demonstrate that the \texttt{TFAW}-corrected light curves of two confirmed exoplanets, K2-44 b and K2-298 b, with high and low SNRs, yield better MCMC posterior distributions thanks to the lower noise contribution. In addition, \texttt{TFAW} yields transit parameters compatible with the cataloged ones but returns smaller uncertainties and narrows the credibility intervals. The \texttt{TFAW} improvement in the photometric performance, transit detection efficiency, and planetary characterization over K2 data can be translated to other running missions, such as \emph{TESS} and  \emph{CHEOPS} \citep{Broeg2013}.

We report the discovery of two statistically validated Earth-sized planets around dwarf stars, EPIC 201170410 and EPIC 201757695, using \texttt{TFAW}-corrected light curves from the \texttt{EVEREST 2.0} database for K2 observing campaign 1. While their small transit depths might not have been detectable and correctly characterized by other algorithms, the combination of the increased photometric precision achieved with \texttt{TFAW}, together with \texttt{TLS} improved capabilities to detect smaller planets identified them as transit candidates. We use a rigorous vetting procedure, the \texttt{vespa} software, the \texttt{EVEREST 2.0} validation, independent photometry and high angular resolution images to statistically validate these candidates. The MCMC characterization of the \texttt{TFAW}-corrected light curves of these candidates reveals that EPIC 201170410.02 is the 10-th smallest planet in K2-C1 and the 44-th in all K2 campaigns; whereas, EPIC 201757695.02 is the 9-th smallest candidate in K2-C1 and the 39-th of all the K2 mission candidates. Work is still in process to fully automate the vetting and FPP computation. A full list of statistically validated candidates for K2-C1 and all other K2 observing campaigns will be presented in a forthcoming paper.

\section*{Acknowledgements}
This research has made use of the NASA Exoplanet Archive, which is operated by the California Institute of Technology, under contract with the National Aeronautics and Space Administration under the Exoplanet Exploration Program. This work made use of NASA ADS Bibliographic Services. This research has made use of \texttt{Aladin} sky atlas developed at CDS, Strasbourg Observatory, France. This work has made use of data from the European Space Agency (ESA) mission Gaia. DdS acknowledges funding support from RACAB. DdS and OF also acknowledge support by the Spanish 
Ministerio de Econom\'ia, Industria y Competitividad (MINEICO/FEDER, UE) under grants AYA2016-76012-C3-1-P, and 
MDM-2014-0369 of ICCUB (Unidad de Excelencia 'Mar\'ia de Maeztu'). DdS and OF acknowledge the support by the Spanish  Ministerio de Ciencia e Innovaci\'{o}n (MICINN) under grant PID2019-105510GB-C31 and through the ``Center of Excellence Mar\'{i}a de Maeztu 2020-2023'' award to the ICCUB (CEX2019-000918-M).

\section*{Data availability}
The data underlying this article will be shared on reasonable request to the corresponding author.



\bibliographystyle{mnras}
\bibliography{mnras} 

\bsp	
\label{lastpage}
\end{document}